\documentclass[modern,manuscript]{aastex61}

\usepackage{rotating}
\usepackage{mathtools}
\usepackage[graphicx]{realboxes}

\newcommand{\lalpha}{Ly$\alpha$}

\newcommand{\pimenc}{$\pi$ Men c}

\newcommand{\pimen}{$\pi$ Men}

\newcommand{\htwo}{H$_2$}
\newcommand{\htwoo}{H$_2$O}
\newcommand{\otwo}{O$_2$}

\newcommand{\hthreep}{H$_3^+$}
\newcommand{\hp}{H$^+$}

\newcommand{\op}{O$^+$}

\newcommand{\chfour}{CH$_4$}
\newcommand{\nhthree}{NH$_3$}
\newcommand{\cotwo}{CO$_2$}

\newcommand{\cp}{C$^+$}

\newcommand{\hi}{H {\sc i}}
\newcommand{\oi}{O {\sc i}}
\newcommand{\cii}{C {\sc ii}}
\newcommand{\ciii}{C {\sc iii}}

\accepted{January 2021}
\submitjournal{ApJL}
\shorttitle{{\pimenc} and the characterization of small exoplanets}
\shortauthors{Garc\'ia Mu\~noz et al.}

\begin{document}

\title{\textbf{A heavy molecular weight atmosphere for the super-Earth {\pimenc}}}

\correspondingauthor{Antonio Garc\'ia Mu\~noz}
\email{garciamunoz@astro.physik.tu-berlin.de, tonhingm@gmail.com}

\author{A. Garc\'ia Mu\~noz}
\affiliation{Zentrum f\"ur Astronomie und Astrophysik, Technische Universit\"at Berlin, 
Hardenbergstrasse 36, D-10623, Berlin, Germany}

\author{L. Fossati}
\affiliation{Space Research Institute, Austrian Academy of Sciences,
Schmiedlstrasse 6, A-8042 Graz, Austria}

\author{A. Youngblood}
\affiliation{
Laboratory for Atmospheric and Space Physics, 1234 Innovation Drive, Boulder, CO 80303, USA}

\author{N. Nettelmann}
\affiliation{
Deutsches Zentrum f\"ur Luft- und Raumfahrt, Rutherfordstrasse, 2, 
Institut f\"ur Planetenforschung, D-12489 Berlin, Germany
}

\author{D. Gandolfi}
\affiliation{
Dipartimento di Fisica, Universit\`a degli Studi di Torino, 
via Pietro Giuria 1, I-10125, Torino, Italy
}

\author{J. Cabrera}
\affiliation{
Deutsches Zentrum f\"ur Luft- und Raumfahrt, 
Institut f\"ur Planetenforschung, D-12489 Berlin, Germany
}

\author{H. Rauer}
\affiliation{
Deutsches Zentrum f\"ur Luft- und Raumfahrt, 
Institut f\"ur Planetenforschung, D-12489 Berlin, Germany
}
\affiliation{
Institute of Geological Sciences, Freie Universit\"at Berlin, 
Malteserstrasse 74-100, D-12249, Berlin, Germany  
}
\affiliation{Zentrum f\"ur Astronomie und Astrophysik, Technische Universit\"at Berlin, 
Hardenbergstrasse 36, D-10623, Berlin, Germany}

\begin{abstract}
\newpage

Strongly irradiated exoplanets develop extended atmospheres that can be utilized to probe the deeper planet layers.
This connection is particularly useful in the study of small exoplanets,  
whose bulk atmospheres are challenging to characterize directly. 
Here we report the 3.4-sigma detection of {\cii} ions during a 
single transit of the super-Earth
{\pimenc} in front of its Sun-like host star. 
The transit depth and Doppler velocities are consistent with the ions filling the planet's Roche lobe 
and moving preferentially away from the star, 
an indication that they are escaping the planet. 
We argue that {\pimenc} possesses a thick atmosphere with abundant heavy volatiles 
($\gtrsim$50{\%} by mass of atmosphere) but that needs not be carbon rich. 
Our reasoning relies upon cumulative evidence from the  
reported {\cii} detection, the non-detection of {\hi} atoms in a past transit,   
modeling of the planet's interior and the assumption that the atmosphere, 
having survived the most active phases of its Sun-like host star, 
will survive another 0.2--2 Gyr. 
Depending on the current mass of atmosphere, {\pimenc} may still 
transition into a bare rocky core.
Our findings confirm the hypothesized compositional diversity of small 
exoplanets,  
and represent a milestone towards understanding the planets' formation and evolution paths
through the investigation of their extended atmospheres. 

\end{abstract}

\keywords{...}

\section{Introduction} \label{sec:intro}

Small exoplanets of sizes between Earth and Neptune are ubiquitous in the galaxy \citep{batalha2014}, 
yet intriguingly absent in our Solar System. 
Even when their masses and radii are accurately known  
\citep{fultonetal2017}, 
little can be confidently stated about their bulk compositions 
\citep{seageretal2007,valenciaetal2010,valenciaetal2013,rogersseager2010,nettelmannetal2011} 
or the processes through which they form and evolve. 
{\pimenc} is a close-in transiting super-Earth \citep{gandolfietal2018, huangetal2018} 
(mass $M_{\rm{p}}$/$M_{\Earth}$=4.52$\pm$0.81; radius $R_{\rm{p}}$/$R_{\Earth}$=2.06$\pm$0.03; 
orbital distance 0.06702$\pm$0.00109 AU)
expected to develop an extended atmosphere under the 
significant XUV (=X-ray + Extreme Ultraviolet) stellar radiation that it receives \citep{kingetal2019, garciamunozetal2020}, 
and is thus an ideal target to investigate the composition of small exoplanets.
Orbiting a Sun-like star, its study also potentially conveys insight into
our own Solar System. 
\\

Whereas planets with 
radii $R_{\rm{p}}/R_{\Earth}$$<$1.6 are likely rocky (composed of iron and silicates)
and those with $R_{\rm{p}}/R_{\Earth}$$>$3 are expected to have a non-negligible amount of 
light volatiles ({\htwo}/He) \citep{otegietal2020}, 
{\pimenc}'s bulk density is consistent with an atmosphere that contains an admixture of light and 
heavy  (e.g. {\htwoo}, {\cotwo}, {\chfour}, {\nhthree}) volatiles 
\citep{rogers2015,otegietal2020}, 
which raises the interesting possibility that it may not be {\htwo}/He-dominated.
Compositional diversity is indeed predicted by theory \citep{
fortneyetal2013,mordasinietal2015}, and supported by the scatter in the
$M_{\rm{p}}$--$R_{\rm{p}}$ statistics of known exoplanets \citep{hatzesrauer2015,otegietal2020}. 
Disentangling the composition of selected small exoplanets
is the key next step, and calls for a multiple line of evidence 
approach that goes beyond $M_{\rm{p}}$ and $R_{\rm{p}}$ measurements. 
\\

Transmission spectroscopy at visible-infrared wavelengths provides 
additional insight when gas absorption bands are revealed \citep{bennekeseager2012}.
However, when the measured spectrum is featureless \citep{guoetal2020}, 
it is difficult to discriminate between 
atmospheres enshrouded by high-altitude clouds and atmospheres with abundant heavy volatiles.
Furthermore, the precision required for visible-infrared spectroscopy approaches the limit of 
current and upcoming telescopes when 
the band-to-continuum contrast drops below $\sim$20 parts per million (as for {\pimenc}).
Here, we alternatively constrain {\pimenc}'s bulk composition with 
far-ultraviolet (FUV) transmission spectroscopy of selected atoms in its extended atmosphere   
complemented with modeling of its interior structure and atmospheric mass loss.
\\

The manuscript is structured as follows. In Section \ref{hstcosobs_cos}, we present 
new FUV transmission spectroscopy measurements of {\pimenc} and argue that the reported dimming 
originates in the planet's atmosphere. 
In Section \ref{extendedatmos_sec}, we describe our upper atmosphere modeling, 
with emphasis on the net mass loss rate, the neutral/ionized state of the escaping hydrogen atoms, 
and their impact on atmospheric stability and detectability of hydrogen atoms. 
In Section \ref{bulkatmos_sec}, we describe our planetary interior modeling, 
which we use to estimate the atmospheric mass for different bulk compositions. 
Lastly, we invoke in Section \ref{discussion_sec} an argument of stability 
that connects the atmospheric mass with a time scale for the planet
to lose it to escape. By requiring that this time scale is not much smaller than the 
system's age, which would suggest fine-tuning in the evolution/current state of the planet, 
we are able to constrain {\pimenc}'s present-day atmospheric composition. 
Appendices \ref{dataanalysis_appendix}--\ref{interior_appendix} provide additional 
technical details.

\section{HST/COS observations \label{hstcosobs_cos}}

We observed one FUV transit of {\pimenc} 
with the Cosmic Origins Spectrograph (COS) aboard the Hubble Space Telescope (HST)
over five consecutive orbits (Program: GO-15699). 
The first two orbits occurred before transit, the third orbit covered the ingress, 
and the last two orbits occurred in transit with respect to the updated ephemeris.
For unknown reasons, the third observation returned no data.
The data were obtained in time-tag mode with the G130M grating centered at 1291 {\AA}. 
Each exposure lasted 3025 s, except the first one that was 
2429 s because of target acquisition prior to the science observation. 
Each spectrum covers the 1135--1429 {\AA} range, with a gap between 1274 and 1291 {\AA}. 
For optimal stability during the observation, 
we adopted one single instrumental Fixed Pattern position (FP-POS=3). 
We downloaded the data from MAST, which were  calibrated and extracted by calcos version 3.3.7\footnote{ 
https://www.stsci.edu/hst/instrumentation/cos/documentation/calcos-release-notes}. 
\\

Each spectrum covers several stellar  
lines of abundant elements (i.e., hydrogen, carbon, nitrogen, oxygen, silicon). 
The lines of the {\oi} 1302--1306 {\AA} triplet, contaminated by geocoronal airglow, 
were treated separately.
The {\hi} {\lalpha} line was also severely affected by geocoronal airglow contamination 
\textcolor{black}{(HST/COS is particularly prone to this problem)}
and gain sag and was not analyzed. 
For each of the other spectral features, 
we integrated the flux in wavelength to obtain a transit light curve for each line 
and ion. 
We either considered each line separately or added together the flux 
from different lines of the same multiplet to increase the S/N 
(i.e., for {\ciii}, N {\sc v}, {\cii}, and Si {\sc iv}). 
In this way, we constructed a transit light curve for each line and ion. 
\\

\begin{figure}[h]
   \centering
   \includegraphics[angle=-90,width=9.cm]{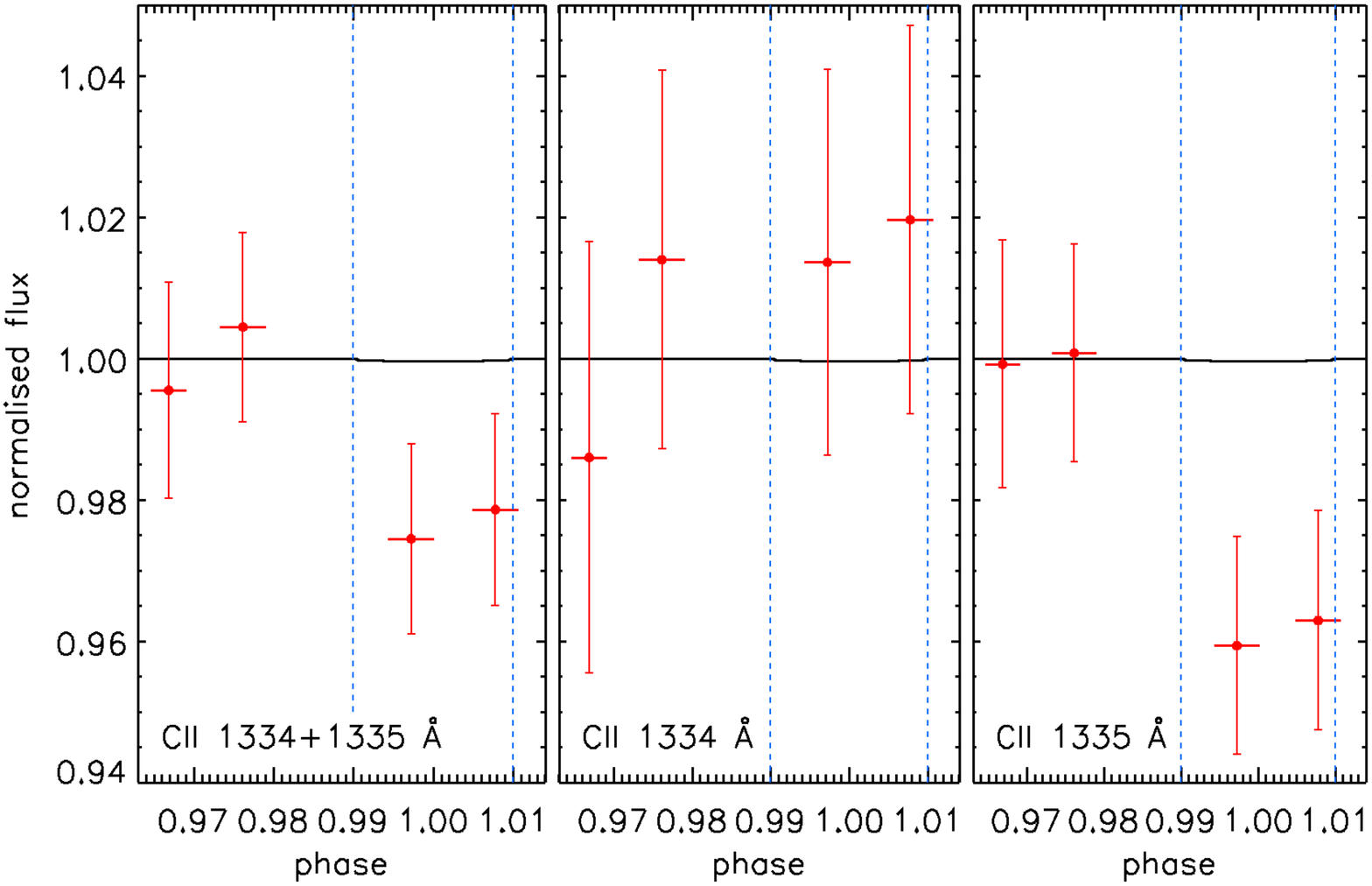}   
   \includegraphics[angle=-90,width=9.cm]{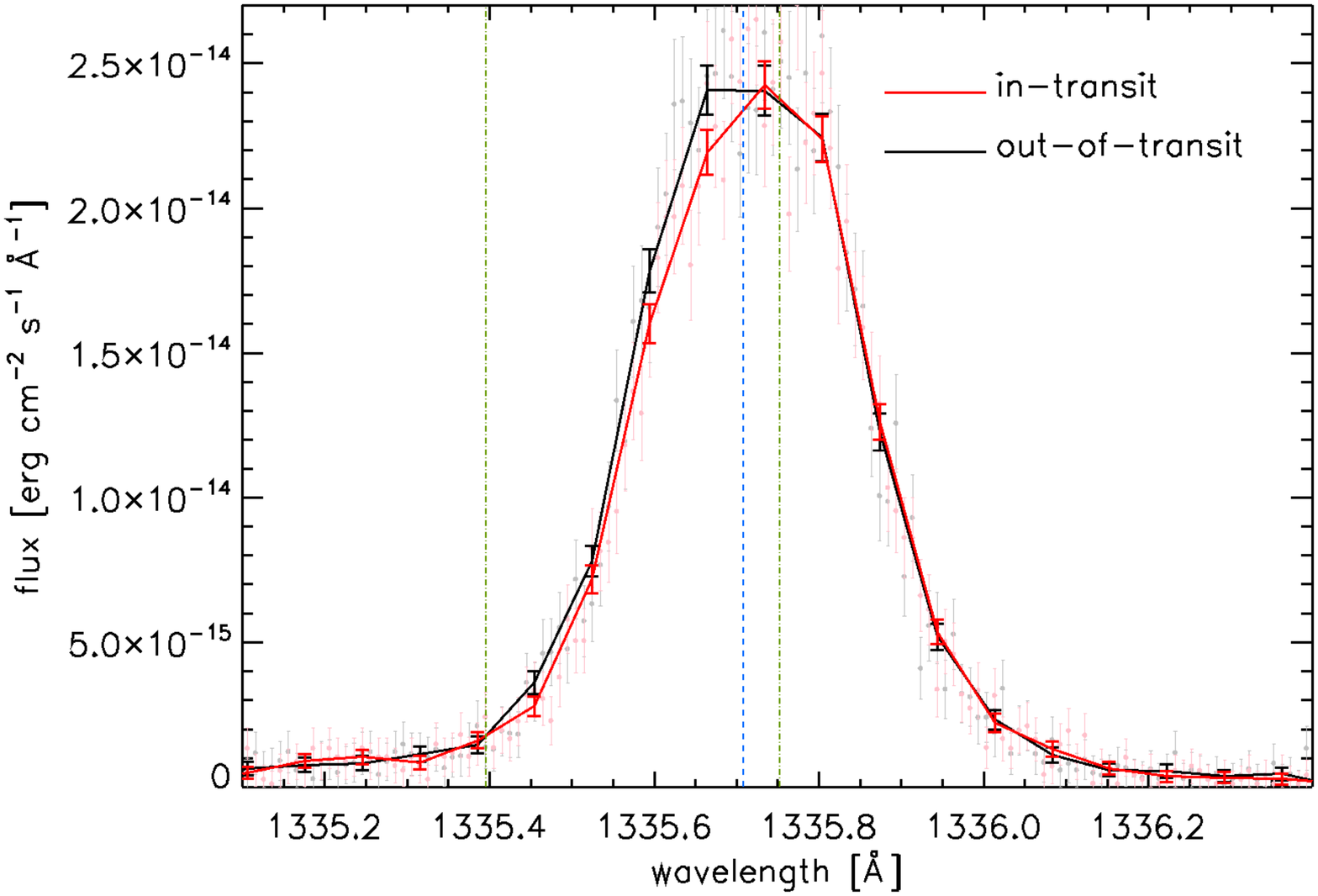}   
   \caption{\label{spectrum_fig}
 \textbf{Top.} 
 Light curves obtained from the COS spectra integrating 
 in wavelength across the entire {\cii} triplet at 1334-1335 {\AA} (left), the {\cii} 1334 {\AA} 
 line (middle), and the {\cii} 1335 {\AA} doublet (right). 
 The horizontal bars indicate the phase range covered by each observation. 
 Each light curve has been normalized to the average flux of the two out-of-transit points. 
 The black line shows the optical transit light curve computed employing published 
 system parameters \citep{gandolfietal2018}. For reference, 
 the blue vertical dashed lines mark the phases of first and last contact.
 \textbf{Bottom.}
 In-transit (red) and out-of-transit (black) spectra around the position of the {\cii} 
 1335 {\AA} doublet. 
 The spectra are rebinned every seven data points for visualization purposes and to match 
 the instrument's spectral resolution, so that each bin corresponds to one resolution element. 
 The gray dots show the unbinned out-of-transit spectrum. The spectra have been brought to 
 the rest frame by accounting for the systemic velocity of the host star. 
 The blue dashed vertical line indicates the position of the main feature composing the {\cii} 1335 {\AA} doublet. 
 The green dash-dotted vertical lines show the integration range considered for measuring 
 the absorption. 
   }
   \end{figure}

Among all light curves, we recorded a significant flux drop during transit only 
for the triplet of {\cii} at 1334--1335 {\AA}.
This feature is composed of a resonance line at 1334.532 {\AA} and a doublet of components at 
1335.663 and 1335.708 {\AA} arising from an excited state (0.007863 eV).
This is why for nearby stars only the bluest line of the triplet is affected by ISM absorption 
(furthermore in the nearby ISM, {\cii} is the dominant C ion \citep{frischslavin2003}). 
The {\cii} 1335-{\AA} doublet is unresolved with COS and the line at 1335.663 {\AA} is about 
10$\times$ weaker than the line at 1335.708 {\AA}, which is the strongest of the triplet. 
The resonance line at 1334.532 {\AA} is intrinsically about 1.8 times weaker than the 
1335.708 {\AA} line. 
In our analysis we ignored the weakest component at 
1335.663 {\AA}. The light curves obtained from splitting the two main lines further 
indicated that the absorption signal was induced by the {\cii} 1335 {\AA} line
(Figure \ref{spectrum_fig}, top). 
Considering only the {\cii} 1335 {\AA} line, the transit depth integrated across the 
whole line (i.e., $\pm$146 km/s from the line center) is 3.9$\pm$1.1{\%}. 
\\

We averaged the in-transit and out-of-transit data separately to obtain one 
master in-transit and one master out-of-transit spectrum and identify the velocities at 
which the absorption in the {\cii} 1335 {\AA} feature occurs. 
Figure \ref{spectrum_fig} (bottom) compares the in- and out-of-transit spectra. 
The {\cii} 1335 {\AA} feature shows in-transit dimming for velocities between $-$70 and $+$10 km/s. 
The corresponding transit depth over this velocity range is 6.8$\pm$2.0{\%} (3.4 sigma detection). 
The corresponding flux drop at the 1334 {\AA} line could not be detected because of interstellar medium absorption (ISM).
We also unsuccessfully looked for absorption in this same velocity range for all the other stellar features
(Table \ref{ExtData_table1}).
\\

Appendix \ref{dataanalysis_appendix} provides further insight into the COS data analysis, 
and discusses the masking  of the {\cii} 1334 {\AA} line by the ISM, 
the unlikeliness that the reported in-transit dimming is caused by random fluctuations
in the stellar line shape and the search for in-transit dimming at the {\oi} 1302--1306 {\AA} triplet.
Our tests suggest that the stellar {\cii} line does not exhibit intrinsic temporal variations
and therefore that the {\cii} 1335 {\AA} line dimming is caused by the planet transit. 
Ideally, future HST/COS observations over one or more transits will confirm the above. 
The confirmation is obviously important for {\pimenc} but also to set useful precedents
in the investigation of other small exoplanets with FUV transmission spectroscopy. 
\\ 

We attribute the dimming of the {\cii} 1335 {\AA} line to absorption by 
{\cii} ions escaping {\pimenc} along with other gases, 
even though the signatures of the other gases are not directly seen in our data. 
Our detection adds to a growing list of exoplanets with extended atmospheres
\citep{vidalmadjaretal2003,
fossatietal2010,benjaffelballester2013,linskyetal2010,ehrenreichetal2015,bourrieretal2018}.
Unlike {\pimenc}, there is strong evidence that these other exoplanets' atmospheres
are {\htwo}/He-dominated.
The transit depth and Doppler velocities reported here are consistent with the {\cii} ions being swept by
the stellar wind into a $\sim$15$R_{\rm{p}}$-wide 
(about the extent of the Roche lobe in the substellar direction, 
\textcolor{black}{and closer to the planet than the interface between the planetary and stellar winds}) 
tail and accelerated to high velocities, a scenario suggested by 
3D models of {\pimenc} and other exoplanets \citep{shaikhislamovetal2020a,shaikhislamovetal2020b}. 
We found by means of a \textcolor{black}{simplified} phenomenological model of {\pimenc}'s tail (Appendix \ref{phenomodel_appendix}) that the 
{\cii} measurements can be explained if the planet loses carbon at a rate $\gtrsim$10$^{8}$ g s$^{-1}$, 
requiring that the atmosphere contains this atom in at least solar abundance.
A tail-like configuration facilitates the detection, but
saturation of the absorption signal impedes setting tighter constraints
on the {\cii} abundance when it becomes supersolar. In summary, 
we cannot \textcolor{black}{yet} conclude 
whether carbon is a major or minor constituent of {\pimenc}'s atmosphere.
Future 3D modeling that incorporates all the relevant physics for the escaping
atmosphere (including the {\cii} ions) and its interaction with the star 
may help discern amongst atmospheric compositions with various carbon abundances.\\

\section{Extended atmosphere \label{extendedatmos_sec}}

A prior observation of {\pimenc} with the HST Space Telescope Imaging Spectrograph (STIS)
revealed no evidence for in-transit  absorption
of the stellar {\lalpha} line \citep{garciamunozetal2020}.
When present, 
absorption in the {\lalpha} wings is primarily caused by energetic neutral atoms (ENAs) 
\citep{holmstrometal2008,tremblinchiang2013}, 
which are fast neutral hydrogen atoms generated when the low-velocity neutral hydrogen 
escaping the planet and the high-velocity protons in the stellar wind exchange charge: 
$$
\rm{H_{planet}} \;(\rm{slow}) + \rm{H_{stellar\;wind}^+} \;(\rm{fast}) \rightarrow 
\rm{H} \;(\rm{fast})\;(\equiv\rm{ENA}) + \rm{H^+} \;(\rm{slow}).
$$
3D modeling shows that if {\pimenc}'s atmosphere is {\htwo}/He-dominated, 
large amounts of ENAs are generated that produce measurable {\lalpha} transit depths
 \citep{shaikhislamovetal2020b}.
Conversely,  
reduced ENA generation occurs if either the flux of stellar wind protons or the flux 
of neutral hydrogen from the planet are weak. The arrangement of {\cii} ions 
into a tail suggests that the stellar wind is not weak, and thus we disfavor the first 
possibility.
A weak neutral hydrogen flux from the planet (the slow component of the above reaction) 
suggests that hydrogen is not the major 
atmospheric constituent or that it ionizes before interacting with 
the stellar wind. 
\\

We investigated {\pimenc}'s extended atmosphere and mass loss 
with a photochemical-hydrodynamic model \citep{garciamunozetal2020}. 
The model takes as input the volume mixing ratio (vmr) 
at the 1 \textcolor{black}{$\mu$bar} pressure level for each species in the chemical network. 
Photo-/thermochemical considerations \citep{mosesetal2013,huseager2014} dictate the most
abundant molecules in the bulk atmosphere given the equilibrium temperature $T_{\rm{eq}}$ and the
fractions of hydrogen, helium, carbon and oxygen nuclei ($Y_{\rm{H}^*}$, $Y_{\rm{He}^*}$, 
$Y_{\rm{C}^*}$, $Y_{\rm{O}^*}$; 
defined as the number of nuclei (symbol $^*$) of each element divided by the total number of nuclei). 
For {\pimenc}'s $T_{\rm{eq}}$$\sim$1150 K, 
the bulk atmosphere composition is dominated by H$_2$ provided that $Y_{\rm{H}^*}$$\sim$1. 
For lower $Y_{\rm{H}^*}$ values, 
other molecules become abundant, such as H$_2$O, CO, CO$_2$ and O$_2$ if 
$Y_{\rm{C}^*}$/$Y_{\rm{O}^*}$$<<$1, or C$_2$H$_2$ and CO if $Y_{\rm{C}^*}$/$Y_{\rm{O}^*}$$>>$1. 
To identify the dominant gases 
in the bulk atmosphere for given sets of nuclei fractions and specify their 
vmrs at the 1 \textcolor{black}{$\mu$bar} pressure level, 
we used a published study of super-Earths and 
sub-Neptunes \citep{huseager2014} (in particular their Figure 7).
We considered bulk compositions in which hydrogen nuclei dominate, 
and compositions in which carbon and oxygen nuclei are also abundant and combine into 
various molecules. 
Table \ref{summaryescape_table} of Appendix \ref{hydromodel_appendix} summarizes 
the implemented vmrs and other derived information 
for the battery of 30 atmospheric runs \textcolor{black}{(labelled as cases 01--30)} that we performed.
Further details on our extended atmosphere modeling are provided in Appendix C.
For the conditions explored here, 
the minimum and maximum loss rates are $\sim$1 and 3{\%} $M_{\rm{p}}$/Gyr. The loss rates 
depend weakly on composition even though the partitioning into different escaping nuclei can be very different.
\\

Our models indicate that the neutral hydrogen flux \textcolor{black}{(the slow component for ENA generation, see above)}
at a reference location defined by the sonic point 
is $\dot{m}_{\rm{H\;I}}$($@$\textit{Mach}=1)$\sim$5$\times$$10^{9}$ g s$^{-1}$
for {\htwo}/He-dominated atmospheres.
From published 3D models for {\htwo}/He-dominated atmospheres 
incorporating ENAs \citep{shaikhislamovetal2020b,holmstrometal2008}, 
we estimate that a neutral hydrogen flux about $\times$1/4 that value will bring the ENA generation
in line with the non-detection of {\lalpha} absorption. 
Thus, 
we set $\dot{m}_{\rm{H\;I}}$($@$\textit{Mach}=1)$\lesssim$1.25$\times$$10^{9}$ g s$^{-1}$ as the approximate
threshold for bulk compositions 
consistent with insufficient ENA generation and therefore with the 
non-detection of {\lalpha} absorption.
Refining this approximate threshold requires the 3D modeling of {\pimenc}'s atmosphere
for a diversity of bulk compositions, which should be the focus of future
investigations. Although welcome, such refinements will not modify the key 
findings of this work.  
\\

The flux $\dot{m}_{\rm{H\;I}}$($@$\textit{Mach}=1) depends strongly on the 
mass fraction of heavy volatiles in the atmosphere $Z$ (=mass of heavy volatiles 
relative to the mass of all volatiles), especially for $Z$$\gtrsim$0.4 when hydrogen
becomes preferentially ionized due to high temperatures (Figure \ref{panel1d_fig})
\citep[see also Fig. 4 in][]{garciamunozetal2020}. 
Remarkably, $\dot{m}_{\rm{H\;I}}$($@$\textit{Mach}=1) depends weakly on 
the identity of the gases contributing to $Z$, 
and indeed the calculated fluxes are comparable for each subset of atmospheric runs with 
different $Y_{\rm{C}^*}$/$Y_{\rm{O}^*}$ ratios.
We exploit the model-predicted $Z$--$\dot{m}_{\rm{H\;I}}$($@$\textit{Mach}=1) relation 
\textcolor{black}{(purple line in Fig. \ref{Matm_Z_fig}, right axis: top panel for cases 01-06; 
bottom panel for cases 07-12)}
to infer that {\pimenc}'s atmosphere has a high $Z$
($\gtrsim$0.85 based on our approximate threshold for $\dot{m}_{\rm{H\;I}}$($@$\textit{Mach}=1),
although the precise $Z$ is subject to the prescribed threshold), 
as otherwise the HST/STIS observation would have revealed {\lalpha} absorption. 
The trend for $Z$--$\dot{m}_{\rm{H\;I}}$($@$\textit{Mach}=1) in Fig. \ref{Matm_Z_fig}
reflects that the partitioning between neutral and ionized hydrogen varies by a larger factor
than the net mass loss rate (which varies by $\sim$3 for the explored $Z$).
\\

\section{Bulk atmosphere \label{bulkatmos_sec}}

We built interior structure models of {\pimenc} that are consistent with its 
$M_{\rm{p}}$, $R_{\rm{p}}$ and $T_{\rm{eq}}$ using a tested methodology \citep{nettelmannetal2011,poseretal2019}. 
The models are organized into a rocky core of iron and silicates in terrestrial proportions, 
and an atmosphere on top containing {\htwo}/He plus a single heavy volatile 
({\cotwo} or {\htwoo}, thus bracketing a broad range of molecular weights). 
This core composition produces $M_{\rm{p}}$--$R_{\rm{p}}$ curves for atmosphereless objects
consistent with the known exoplanets
that presumably lack a volatile envelope \citep{otegietal2020}. 
When considering {\htwoo} as the dominant heavy volatile, it is assumed that 
another carbon-bearing molecule present in trace amounts carries the carbon detected in
the HST/COS data. We assume that all gases remain well mixed, 
\textcolor{black}{which is justifiable for reasonable values of the eddy diffusion
coefficient in the atmosphere and the mass loss rates estimated here \citep{garciamunozetal2020}}. 
The {\htwo}/He mass fraction is kept constant to the protosolar value, 
but $Z$ is varied to explore atmospheres with different abundances of heavy volatiles. 
\textcolor{black}{We can thus pair the interior structure models and the upper atmosphere
models on the basis of their corresponding $Z$ and the dominating molecules (or more generally the
$Y_{\rm{C}^*}$/$Y_{\rm{O}^*}$ ratio in the upper atmosphere). }
\\

The models consider a \textcolor{black}{present-day} intrinsic temperature ($T_{\rm{int}}$, which specifies the heat flux from the
interior through Stefan-Boltzmann's law $\sigma_{\rm{B}}$$T_{\rm{int}}^4$) and an extra opacity
as adjustable parameters. 
We find that for $T_{\rm{int}}$$>$100 K all plausible atmospheres are relatively
light and would reach the current state within a time much shorter than the system's age 
(5.2$\pm$1.1 Gyr) and then would continue cooling and contracting
\textcolor{black}{(Fig. \ref{planetinterior_fig}, top left)}. 
We consider this temperature to be a conservative upper bound. 
Remarkably, for all other parameters being the same, 
a higher $T_{\rm{int}}$ translates into a \textcolor{black}{less massive} atmosphere that is easier to lose.
\textcolor{black}{Indeed, increasing $T_{\rm{int}}$ 
causes larger scale heights and in turn larger atmospheric volumes for the same atmospheric mass.}
The best match betwen \textcolor{black}{evolution} models \textcolor{black}{that include stellar irradiation 
as the sole external energy source} and the measured $R_{\rm{p}}$ after cooling for 3 Gyr
occurs for $T_{\rm{int}}$=44--52 K \textcolor{black}{(Fig. \ref{planetinterior_fig}, top left)}, 
in which case the atmosphere has reached equilibrium with the incident irradiation. 
In what follows, we focus on the range $T_{\rm{int}}$=44--100 K. 
Key model outputs are the atmosphere and core masses 
($M_{\rm{atm}}$+$M_{\rm{core}}$=$M_{\rm{p}}$) and the core radius ($R_{\rm{core}}$$\le$$R_{\rm{p}}$).
These quantities ($M_{\rm{atm}}$, $R_{\rm{core}}$) are determined with no prior assumption 
on their values by iteratively solving 
the interior structure equations so that upon convergence the model complies with the specified planet mass and radius 
constraints.
For reference, the core size turns out to be always $R_{\rm{core}}$/$R_{\Earth}$$\sim$1.6 for $Z$$\ll$1, 
but can be $\sim$1.4 for $Z$({\htwoo})=0.9 and 
$\sim$1 for $Z$({\cotwo})=0.9.
Appendix \ref{interior_appendix} provides additional insight into the interior structure model.
\\

{\htwo}/He-dominated atmospheres ($Z$$\ll$1) \textcolor{black}{are more} voluminous but 
\textcolor{black}{overall} contribute little mass (Figure \ref{Matm_Z_fig}, left). 
For example, $M_{\rm{atm}}$/$M_{\rm{p}}$$<$2$\times$10$^{-3}$ for $Z$({\htwoo}) or $Z${(\cotwo})=0.3.
In turn, an atmosphere with abundant heavy volatiles must be massive
to compensate for its reduced scale height. Thus, 
$M_{\rm{atm}}$/$M_{\rm{p}}$$\sim$7$\times$10$^{-2}$ for $Z$({\htwoo})=0.8, 
and as high as $\sim$0.2 for $Z$({\cotwo})=0.8.
However, not every atmospheric composition consistent with the interior models 
is stable over long timescales.
We estimated the mass of atmosphere that is lost over a range of times
$t_{\rm{XUV}}$  as $\dot{m}\;t_{\rm{XUV}}$/$M_{\rm{p}}$
(dashed lines, Figure \ref{Matm_Z_fig}), where $\dot{m}$ is the
loss rate predicted by our photochemical-hydrodynamic models and that varies as the stellar XUV luminosity and the 
planet orbital distance evolve. We incorporate these effects into $t_{\rm{XUV}}$, which must
be viewed as an equivalent time based on the current stellar luminosity and orbit.

\section{Atmospheric stability and composition \label{discussion_sec}}

Using arguments of atmospheric stability to constrain {\pimenc}'s interior 
requires an appropriate timescale over which the atmosphere will survive. 
We \textcolor{black}{first} adopted a survival time of 2 Gyr, which is a moderate fraction of the 
time left before the star exits the main sequence ($\sim$5 Gyr) and assumes that
{\pimenc} is not near catastrophic mass loss. 
This choice implicitly assumes that if the atmosphere survived the 
$\times$100--1000 enhancement in XUV luminosity experienced in the early 
life of its host star, then its end might not be imminent.
Under this hypothesis, we infer (Figure \ref{Matm_Z_fig})
that $Z$({\htwoo})$\ge$0.73 and $Z$({\cotwo})$\ge$0.65. 
These are conservative bounds based on the uppermost sets of 
interior model curves for each heavy volatile, 
and correspond to volume mixing ratios vmr({\htwoo})$\ge$0.26 (molecular weight 
$\mu$$\ge$6.4 g mol$^{-1}$) and vmr({\cotwo})$\ge$0.09 ($\mu$$\ge$6 g mol$^{-1}$).  
A ten-fold shorter survival time results in 
$Z$({\htwoo})$\ge$0.50 (vmr({\htwoo})$\ge$0.11; $\mu$$\ge$4.1 g mol$^{-1}$) and 
$Z$({\cotwo})$\ge$0.45 (vmr({\cotwo})$\ge$0.04; $\mu$$\ge$4 g mol$^{-1}$). 
\\

The inferred heavy mass fractions $Z$ are approximately consistent with the non-detection 
of {\lalpha} absorption, which renders independent support to our findings. It ultimately confirms 
that a thick atmosphere with more than half its mass in 
heavy volatiles is a realistic scenario for {\pimenc}. 
For comparison, maximum values of vmr({\htwoo})$\sim$0.09--0.15 have been inferred
from infrared spectroscopy for the only other exoplanet with $R_{\rm{p}}$$<$3$R_{\Earth}$ 
at which {\htwoo} has been detected \citep{bennekeetal2019,tsiarasetal2019,madhusudhanetal2020}. 
Thus, {\pimenc} becomes the exoplanet with the highest abundance of heavy volatiles known to
date, and its case suggests that even higher abundances might be expected 
for other small exoplanets. 
It is uncertain how the planet acquired such a heavy atmosphere, although high-$Z$ 
atmospheres are natural outcomes of formation models \citep{fortneyetal2013}.
Assuming that water is the dominant heavy volatile, it is plausible that {\pimenc} might have formed beyond the snow line
and reached its current orbit following high-eccentricity migration and tidal circularization. 
The idea is supported by {\pimenc} being on a misaligned orbit with respect 
to the stellar spin axis \citep{kunovachodzicetal2020}
and the fact that the system contains a far-out gas giant on an eccentric, inclined orbit 
\citep{damassoetal2020,derosaetal2020,xuanwyatt2020}.
\\

{\pimenc} lies near the so-called radius gap \citep{owenwu2013,fultonetal2017}
that separates the population of planets 
that presumably lost their volatiles through atmospheric escape 
(peak at $R_{\rm{p}}$/$R_{\Earth}$$\sim$1.5)
from those that were able to retain them (peak at $R_{\rm{p}}$/$R_{\Earth}$$\sim$2.5). 
The planet may still lose up to 10{\%} of its mass
in the future 5 Gyr if it remains on its current (and stable) orbit \citep{derosaetal2020,xuanwyatt2020}. 
This is more than what the planetary interior models predict for $M_{\rm{atm}}$/$M_{\rm{p}}$ 
for some plausible atmospheric configurations.
It is thus likely that we are witnessing {\pimenc} while crossing the radius gap. 
Indeed, Figure \ref{Matm_Z_fig} suggests that this will happen unless 
the actual $Z$({\htwoo})$\gtrsim$0.85 or $Z$({\cotwo})$\gtrsim$0.80.
In that event, and because $M_{\rm{atm}}$/$M_{\rm{p}}$$<<$1 for such unstable 
configurations, the remnant core will collapse onto the empirical 
$M_{\rm{p}}$--$R_{\rm{p}}$ curve for atmosphereless objects.  
\\

\newpage

\begin{table}[h]
\caption{
In-transit absorption measured over the velocity range $-$70 to $+$10 km/s
for various stellar emission lines in the COS data. 
}             
\label{ExtData_table1}      
\centering                          
\begin{tabular}{c c c c}        
\\
\hline                 
Ion & Wavelength  & In-transit        & Statistical \\
    &  [{\AA}]    & absorption [$\%$] & significance \\
\hline    
{\cii} & 1335.7    & 6.76$\pm$2.00 & 3.39 \\
\hline
{\oi}  & 1302.168  & $-$3.65$\pm$40.65 & 0.09 \\ 
{\oi}  & 1304.858  & $-$8.33$\pm$6.32  & 1.32 \\
{\oi}  & 1306.029  & 5.17$\pm$3.66  & 1.41 \\
\hline
{Si {\sc iii}}  & 1206.5  & 1.49$\pm$2.01  & 0.74 \\
{N {\sc v}}  & 1238.821  & $-$0.82$\pm$5.37  & 0.15 \\
{N {\sc v}}  & 1242.804  & $-$17.26$\pm$9.97  & 1.73 \\
{Si {\sc ii}}  & 1265.002  & 3.38$\pm$4.94  & 0.68 \\

Cl {\sc i}  & 1351.656  & 6.12$\pm$7.65  & 0.80 \\

{\oi}  & 1355.598  & 7.36$\pm$5.87  & 1.25 \\

{Si {\sc iv}}  & 1393.755  & 8.41$\pm$3.09  & 2.72 \\

{Si {\sc iv}}  & 1402.770  & $-$0.53$\pm$5.07  & 0.11 \\

\hline

All ions excluding       &   & 2.67$\pm$1.27 & 2.10 \\
{\cii} and {\oi} triplet &   &   & \\

\hline                                   
\end{tabular}
\end{table}

\newpage

\newpage

   \begin{figure}[h]
   \centering
\includegraphics[angle=0,width=8.cm]{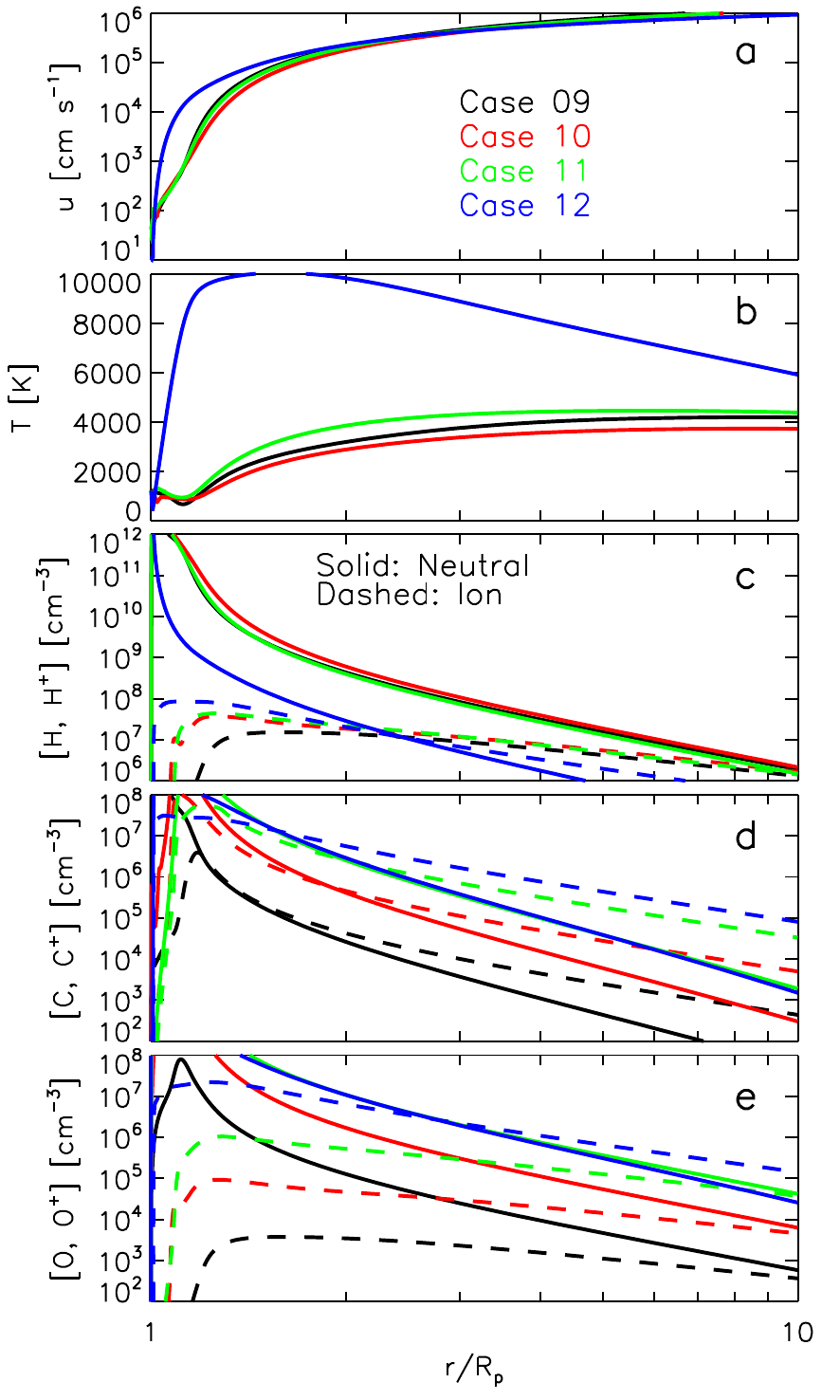}
   \caption{\label{panel1d_fig}
   Photochemical-hydrodynamic solutions for selected cases from Table \ref{summaryescape_table}. 
   \textcolor{black}{These cases cover a broad range of heavy volatile abundances from $Z$=4.8$\times$10$^{-3}$ to 0.9.
   \textbf{a}) Velocity profiles. Most of the gas acceleration occurs below $r$/$R_{\rm{p}}$$\sim$3, at which location the gas reaches velocities of a few km/s. 
   \textbf{b}) Temperature profiles. Typically, the temperatures remain well below 4000-5000 K except for the higher $Z$ that 
   it reaches up to 10000 K. 
   This trend is seen over the entire set of 30 atmospheric runs. 
   \textbf{c}) Number density profiles for H and {\hp}. For the larger $Z$, the transition 
   between these two states occurs notably closer to the planet. 
   \textbf{d}) Number density profiles for C and {\cp}. The C atom ionizes closer to the planet
   than the H and O atoms.    
   The {\cp} photoionization lifetime is long enough ($t_{\rm{C\;II}}$$\sim$20 hours) that
   can form a long tail trailing the planet. 
   \textbf{e}) Number density profiles for O and {\op}.
   }   
   }
   \end{figure}

\newpage

   \begin{figure}[h]
   \centering
   \includegraphics[angle=0,width=13.cm]{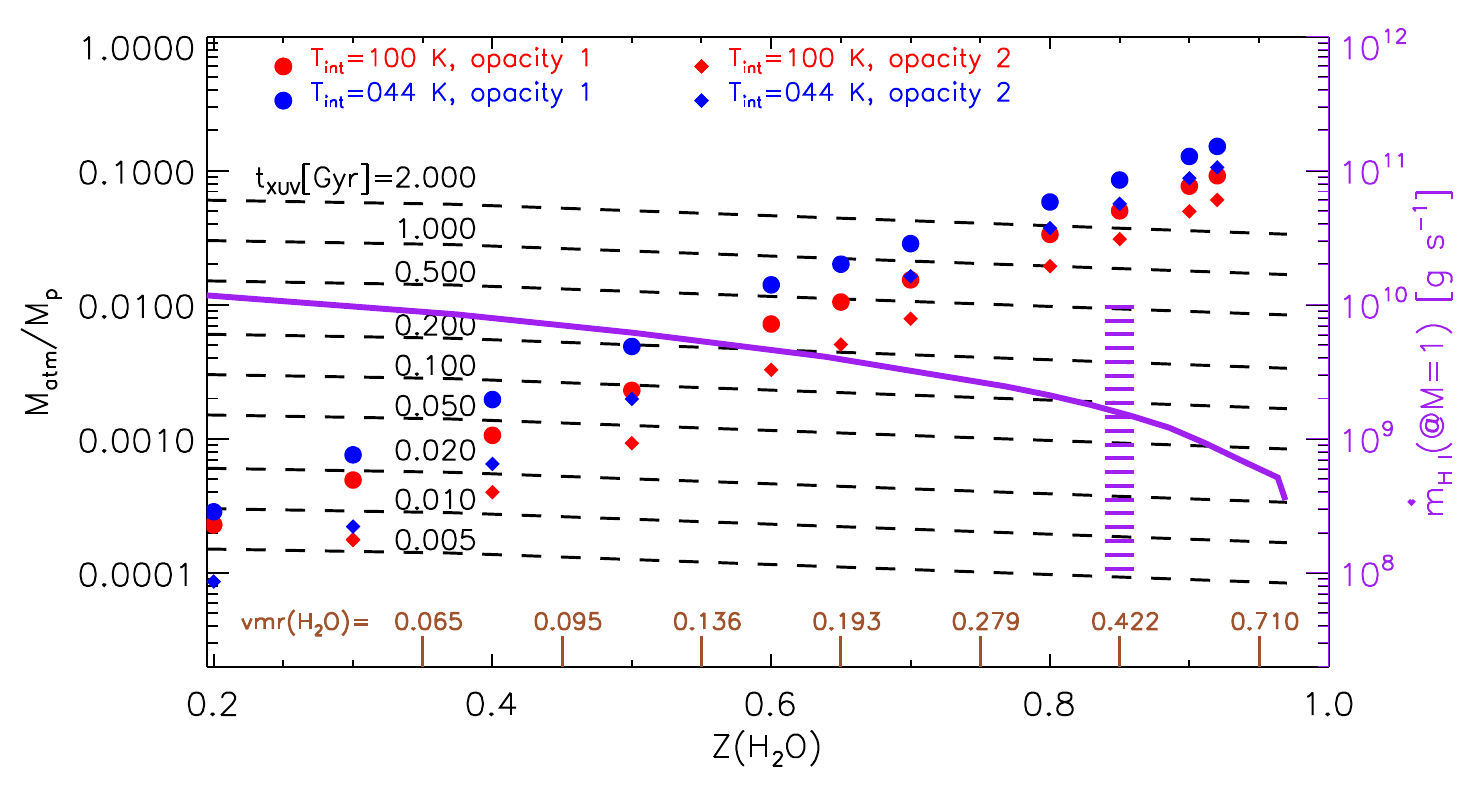}
   \includegraphics[angle=0,width=13.cm]{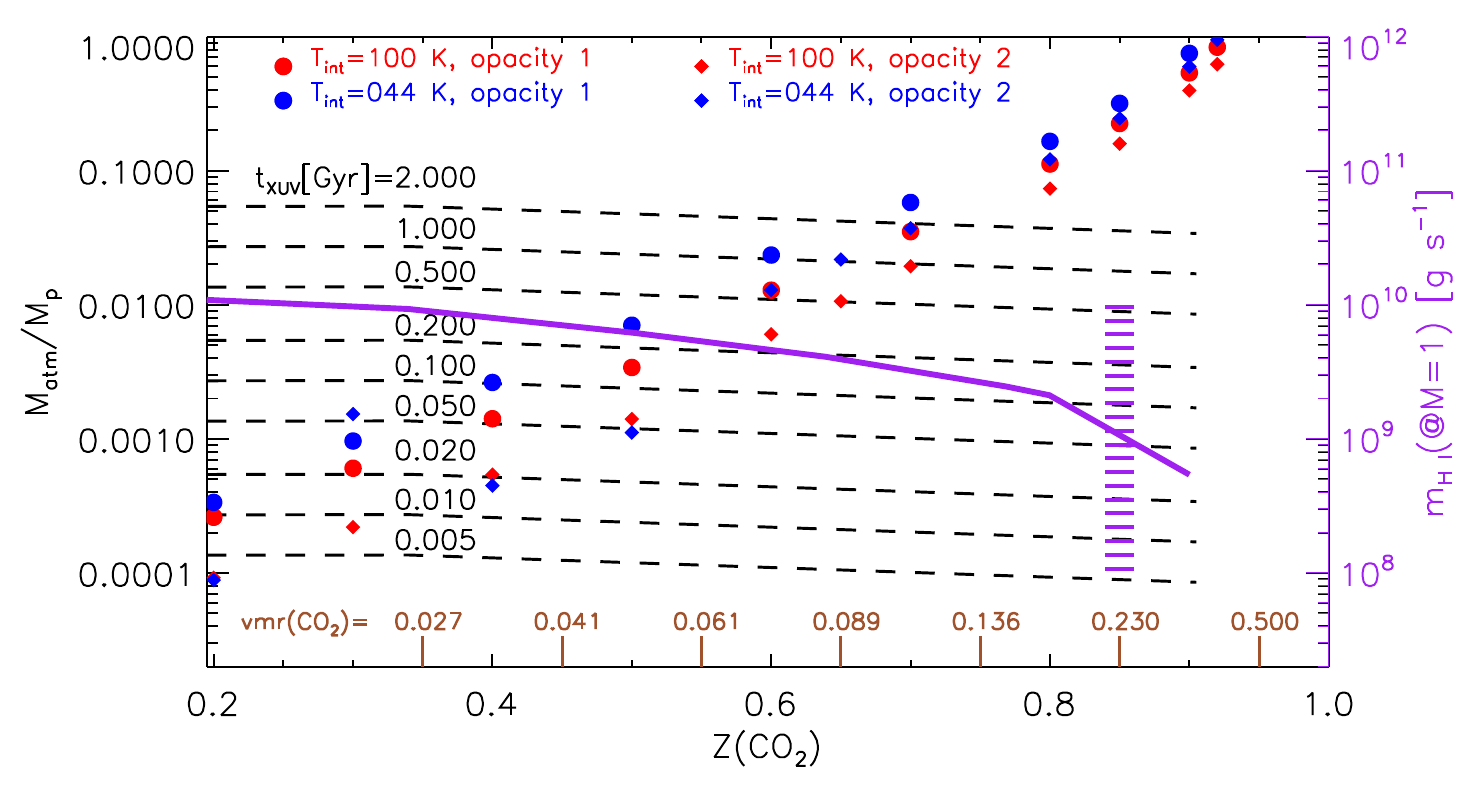}   
   \caption{\label{Matm_Z_fig} 
Atmospheric mass normalized to the measured planet mass as calculated by
internal structure models for various choices of $T_{\rm{int}}$ and gas opacity. 
The modeling assumes {\htwo}/He plus a heavy 
volatile ({\htwoo}, \textbf{top}; {\cotwo}, \textbf{bottom}), and explores
mass fractions of heavy volatiles $Z$ from 0.2 to about 1. For reference, the corresponding
volume mixing ratio of the heavy volatile is also shown (brown). 
The black dashed lines show the mass fraction lost over the indicated times, from 
$t_{\rm{XUV}}$=0.005 to 2 Gyr.  
\textcolor{black}{For the top (bottom) panel, 
we used the mass loss rates quoted in Table \ref{summaryescape_table} for cases 01--06 (07--12);  
the choice between subsets of atmospheric runs with different $Y_{\rm{C}^*}$/$Y_{\rm{O}^*}$ ratios
is not critical though.}
The top panel (purple line, right axis) shows the
neutral hydrogen fluxes from the planet at the sonic point predicted by the
photochemical-hydrodynamic model \textcolor{black}{for cases 01--06}.
\textcolor{black}{Correspondingly, the bottom panel shows the fluxes predicted for cases 07--12.}
\textcolor{black}{High neutral hydrogen fluxes (which occur for smaller $Z$) will 
result in enhanced ENA generation and in turn increased {\lalpha} transit depths.}
The dashed purple pattern 
indicates our approximate threshold ($Z$=0.85) separating atmospheric conditions that 
result in detectable and undetectable {\lalpha} absorption. 
   }
   \end{figure}

\newpage

\acknowledgments

Davide Gandolfi gratefully acknowledges financial support from the CRT foundation under grant No. 2018.2323 "Gaseous
or rocky? Unveiling the nature of small worlds".

\newpage

\clearpage

\appendix

\section{Preparation of the observations and data analysis \label{dataanalysis_appendix}}

We improved the published transit ephemeris \citep{gandolfietal2018}
using TESS data from Sectors 1, 4, 8, 11--13 and the code pyaneti \citep{barraganetal2019}, 
which allows for parameter estimation from posterior distributions. 
\\

\begin{figure}[h]
   \centering
   \includegraphics[angle=-90,width=9.cm]{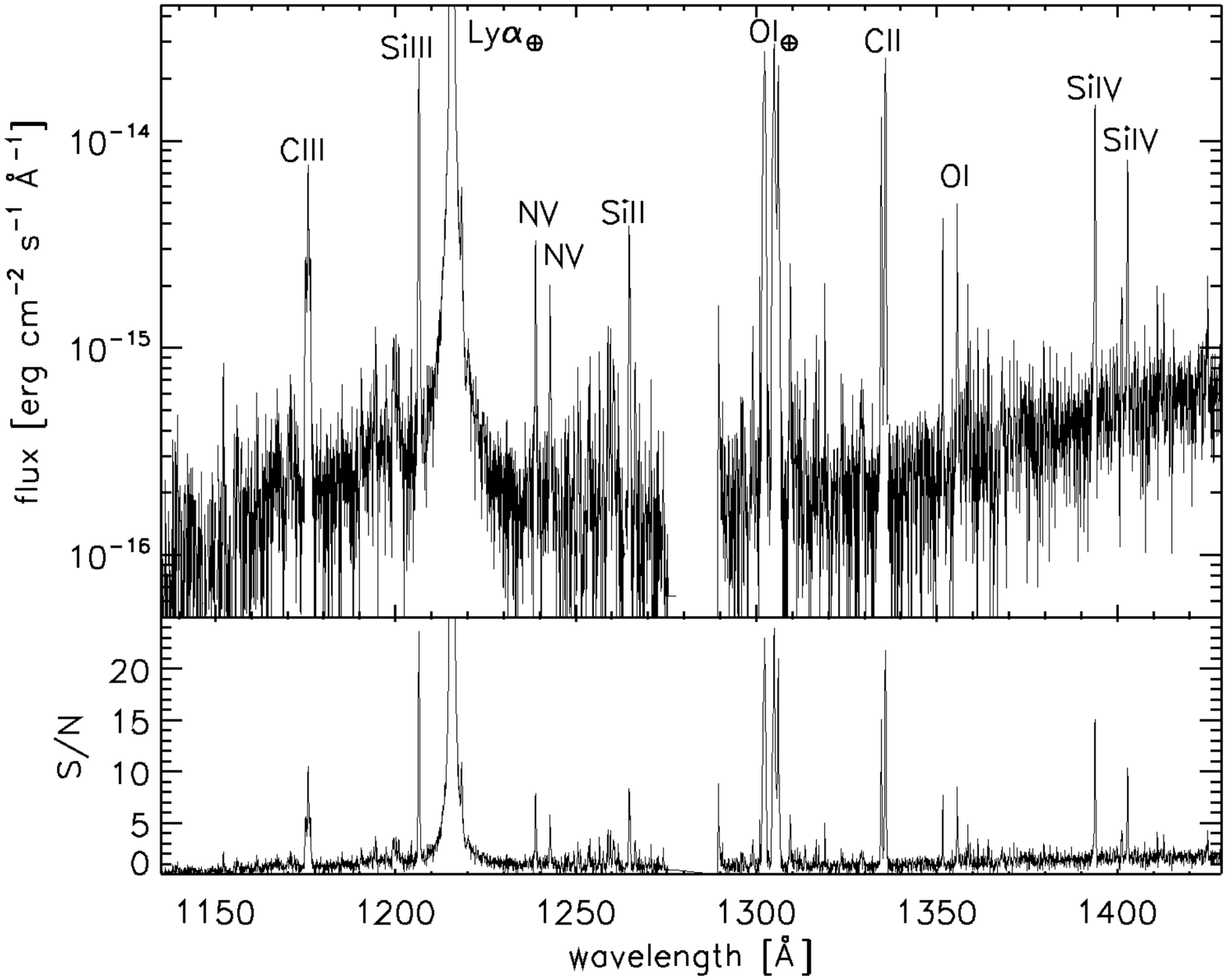}
   \includegraphics[angle=-90,width=11.cm]{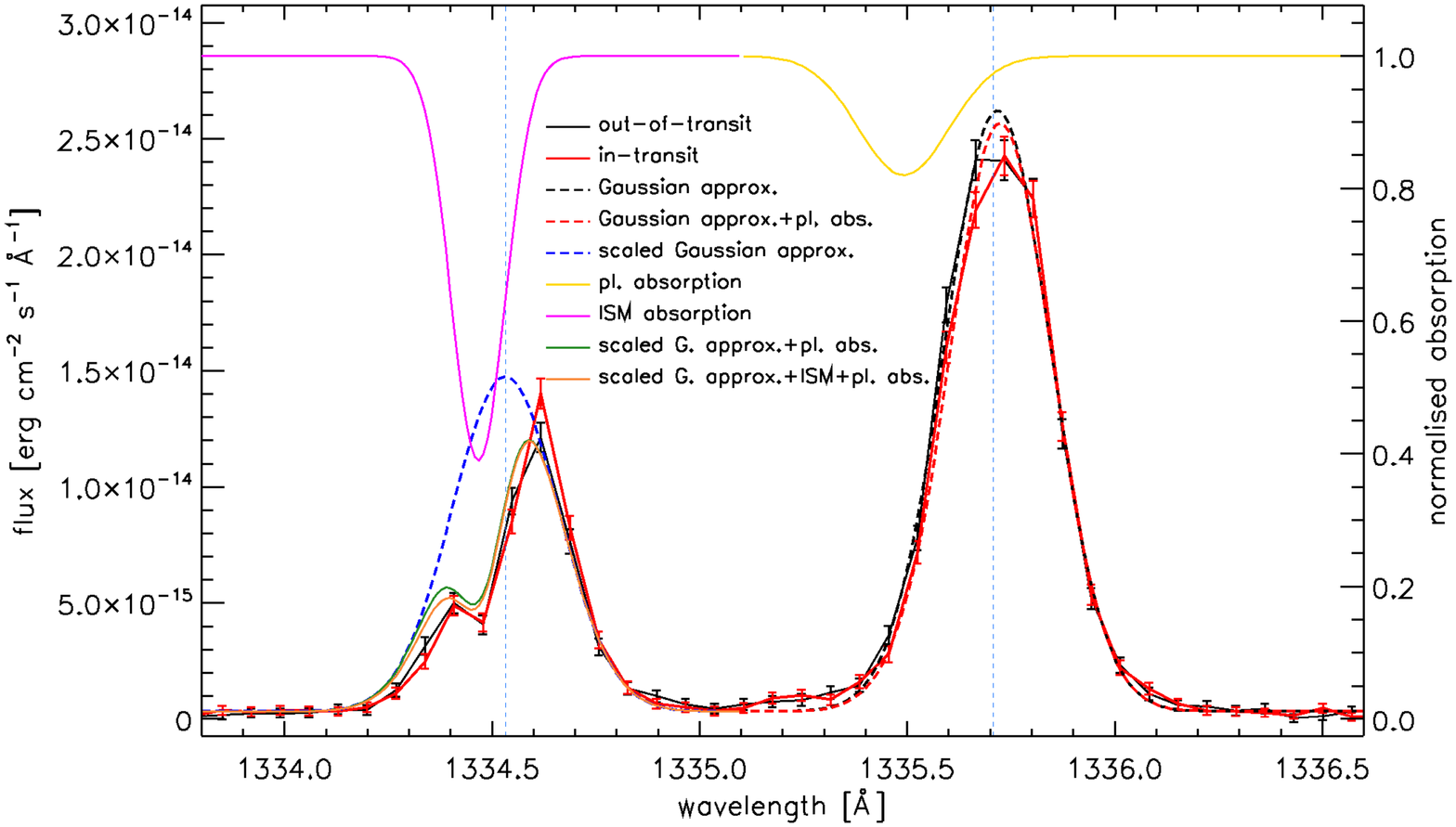}   
   \caption{\label{CIIanalysis_fig}
   \textbf{Top}. Far-ultraviolet spectrum of {\pimen} obtained during the second HST orbit. 
   Major features are marked. The Earth symbol indicates that the {\lalpha} line and the 
   {\oi} triplet are affected by geocoronal airglow. 
   \textbf{Middle}. S/N per pixel of the top panel spectrum.
\textbf{Bottom}.
Rebinned out-of-transit (black) and in-transit (red) spectra of {\pimen} around the {\cii} 
1334/1335 {\AA} triplet. The black dashed line shows the Gaussian fit to the {\cii} 1335 {\AA} 
feature, while the blue dashed line shows the simulated {\cii} 1334 {\AA} line, 
without ISM absorption, obtained scaling the Gaussian fit to the {\cii} 1335 {\AA} feature by the 
ratio of the line oscillator strengths times the statistical weights of the lower states. 
The yellow line shows the fitted planetary absorption 
(right y-axis). The red dashed line shows the Gaussian fit to the {\cii} 1335 {\AA} 
feature multiplied by the simulated planetary absorption. 
The purple solid line shows the simulated ISM absorption (right y-axis), 
while the green solid line shows the simulated {\cii} 1334 {\AA} line profile multiplied 
by the ISM absorption profile. The orange solid line shows the {\cii} 1334 {\AA} line 
profile multiplied by the ISM and planetary absorption profiles. 
The vertical blue dashed lines indicate the positions of the two main components of the 
{\cii} triplet.
}
   \end{figure}

Figure \ref{CIIanalysis_fig} shows the spectrum obtained during the second HST observation (top), 
marking the stellar features at which we looked for absorption in the $-$70 to $+$10 km/s velocity range, 
and the resulting signal-to-noise ratio per spectral bin (middle).
The bottom panel compares the in- and out-of-transit spectra.
\\

Having detected absorption of the {\cii} 1335 {\AA} feature, we looked for a 
similar signal at the {\cii} 1334 {\AA} line, without success.
Next, we show that the absorption signal at 1334 {\AA} is hidden by ISM contamination
(Figure \ref{CIIanalysis_fig}, bottom), which affects the S/N.
We first fitted the {\cii} 1335 {\AA} stellar feature using a Gaussian profile 
and the out-of-transit spectrum (black dashed line) and then employed a further Gaussian profile to 
fit the position and strength of the planetary
absorption on the in-transit spectrum (yellow solid line).
We obtained that the Gaussian profile simulating the planetary absorption lies at a velocity of 
about $-$48.5 km/s with respect to the position of the main {\cii} 1335 {\AA} feature, 
has a normalized amplitude of $\sim$0.18, and a width of $\sim$0.11 {\AA}. 
These fits were performed considering the unbinned spectra. 
We then derived the strength of the {\cii} 1334 {\AA} line, prior to ISM absorption by 
scaling the Gaussian fit to the {\cii} 1335 {\AA} feature by the ratio of the oscillator 
strength times the statistical weight of the two lines (about 1.8; blue dashed line). 
We further simulated the {\cii} ISM absorption profile at the position of the {\cii} 
1334 {\AA} resonance line employing a Voigt profile (purple solid line), 
in which we set the position of the line equal to that obtained 
from the reconstruction of the stellar {\lalpha} line \citep{garciamunozetal2020} 
and a {\cii} ISM column density equal to that of hydrogen 
scaled to the expected ISM carbon-to-hydrogen abundance ratio and ISM C ionization fraction 
\citep{frischslavin2003} (green solid line). 
Figure \ref{CIIanalysis_fig} (bottom) indicates that the simulated profile is a good 
match to the out-of-transit spectrum, particularly accounting for the uncertainties 
involved in inferring the {\cii} ISM column density. Finally, we added to this line the 
planetary absorption obtained from fitting the {\cii} 1335 {\AA} feature, 
rescaled by 1.8 (orange solid line). 
Lastly, Figure \ref{CIIanalysis_fig} (bottom) shows that the difference between the simulated 
{\cii} 1334 {\AA} line profiles before and after adding the planetary absorption is 
significantly smaller than the observational uncertainties and hence undetectable in the data.
In summary, reduced S/N due to ISM contamination hides the planetary absorption signal at 1334 {\AA}. 
\\

To estimate the likelihood that our {\cii} signal is due to intrinsic stellar 
variability, we did a Monte Carlo simulation where we assumed the population of 
intrinsic stellar variability between the mean in-transit and mean out-of-transit spectrum 
is represented by the measured in-transit absorption from each line (Table \ref{ExtData_table1}). 
We excluded from this representative population the {\cii} lines because of the putative 
planetary absorption and the {\oi} triplet because the noise properties and intrinsic 
variability of these airglow-contaminated lines are different. 
For the remaining eight emission lines, we calculate a weighted mean of 
2.67$\pm$1.27{\%} for the in-transit absorption. We made 10$^6$ realizations of nine emission 
lines (the eight 'unaffected' lines of the representative population plus {\cii} 1335.7 {\AA}, 
the line with the 3.4-sigma in-transit absorption) each with in-transit absorption 
randomly drawn from a normal distribution with mean 2.67{\%} and standard deviation 1.27{\%}. 
We find that the likelihood of 1 or more emission lines having an absorption $\ge$6.76{\%} 
due to intrinsic stellar variability as measured by our COS spectra is 2{\%}.
\\

\begin{figure}[h]
   \centering
   \includegraphics[angle=0,width=10.cm]{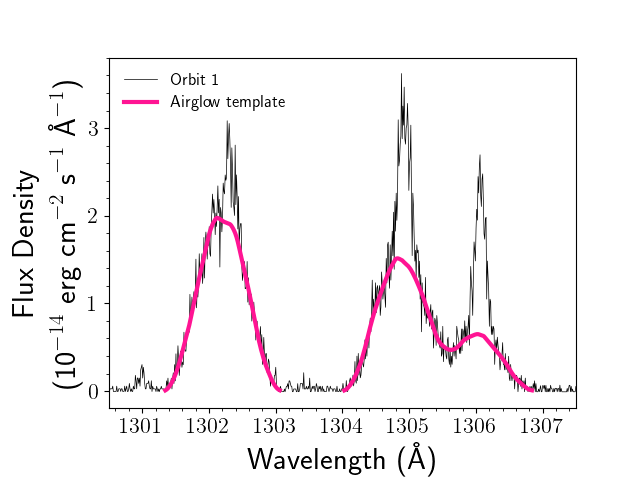}
   \includegraphics[angle=0,width=13.cm]{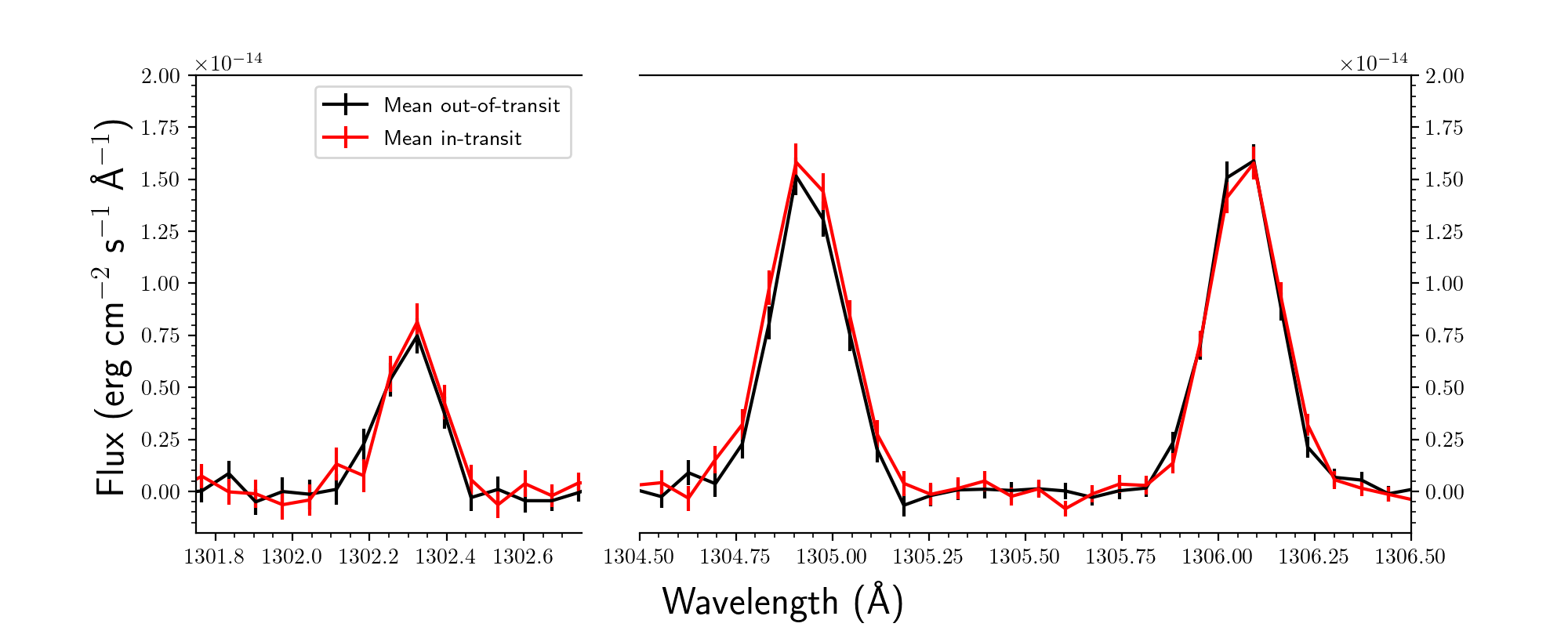}   
   \caption{\label{OIanalysis_fig}
   \textbf{Top}.
Geocoronal airglow template \citep{bourrieretal2018cos} fits to Orbit 1 of our 
unbinned COS spectra. The narrower emission lines superimposed on the broader airglow 
lines have a stellar origin. These stellar lines were masked out from the fit. 
Orbits 2, 4, and 5 exhibit a similarly good fit.   
\textbf{Bottom}. 
   For the {\oi} triplet, mean in-transit (red) and mean out-of-transit (black) spectra
   binned in the same way as the C II spectra after correcting for geocoronal airglow. 
For the line at 1306 {\AA} (the least affected by geocoronal airglow), 
we estimate a transit depth of 5.2$\pm$3.7{\%} (statistically insignificant)
over the velocity range from $-$70 to +10 km/s.
  }
   \end{figure}

\begin{figure}[h]
\centering
\includegraphics[angle=0,width=13.cm]{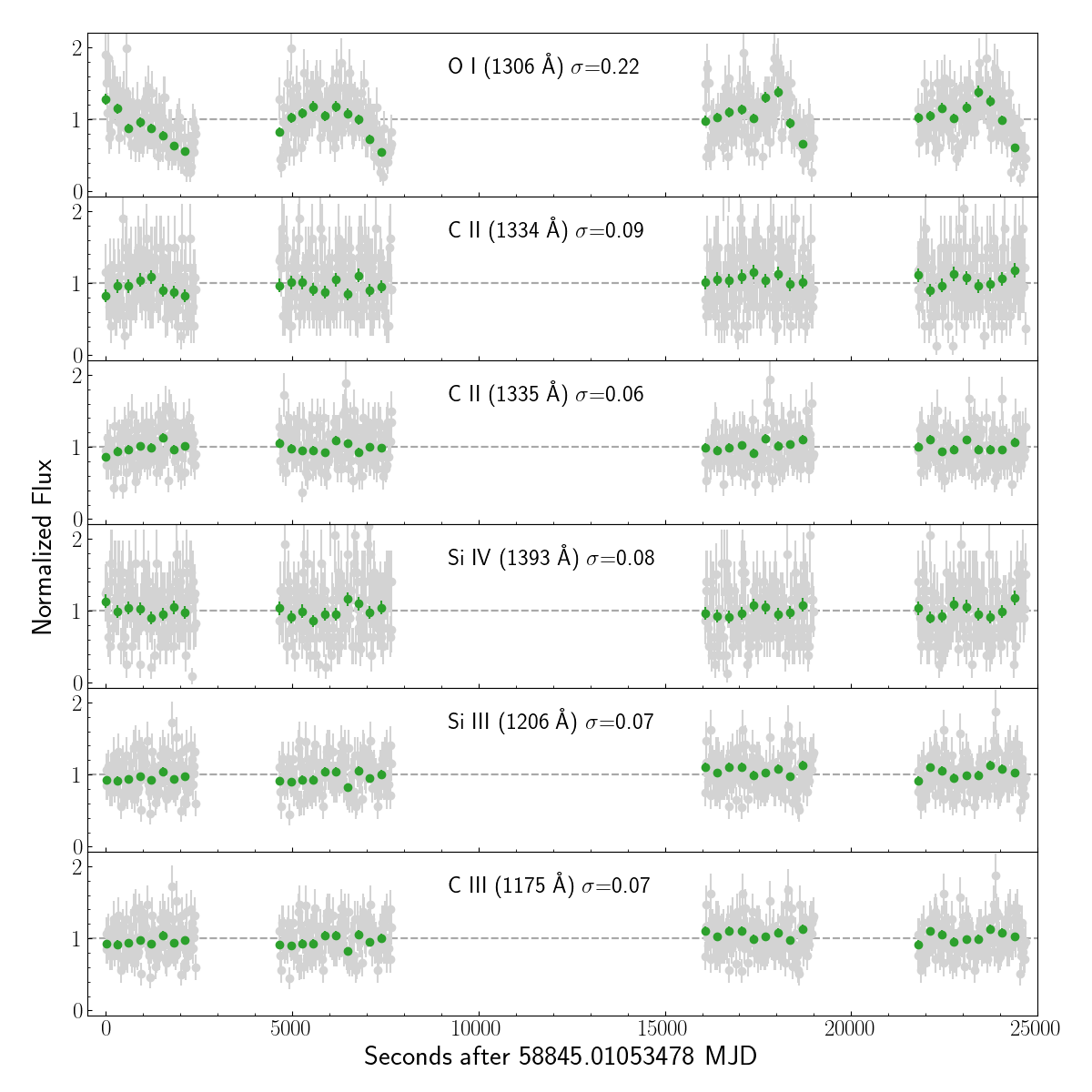}   
\caption{ \label{picture7ay_fig}
Normalized light curves for a selection of the observed emission lines. 
The gray points show 20-s bins and the green points show 300-s bins. 
The standard deviation of the 300-s bin points is printed in each subplot. 
The {\oi}’s light curve is dominated by airglow variability over the course of each orbit. 
No flares or other short term stellar variability are apparent. 
}
\end{figure}

We inspected the {\oi} lines for evidence of planetary absorption. 
The {\oi} triplet at 1302--1306 {\AA} is comprised of three emission lines that share an upper 
energy level. 
{\oi} 1302.168 {\AA} is the resonance line, and is significantly affected by 
ISM absorption. {\oi} 1304.858 and 1306.029 {\AA} each have similar oscillator strengths 
to the resonance line. COS's wide aperture leads to significant contamination of each 
line of the {\oi} triplet by geocoronal airglow. We corrected each orbit's 
pipeline-reduced x1d spectrum's {\oi} emission lines for geocoronal airglow by performing a 
least-squares fit of the airglow templates downloaded from MAST \citep{bourrieretal2018cos} 
to the spectrum (Figure \ref{OIanalysis_fig}, top). 
We created mean in-transit and out-of-transit spectra in the same way as the {\cii} lines, 
and measured absorptions in each of the three lines (shortest wavelength to longest wavelength) 
of $-$3.65$\pm$40.65 {\%}, $-$8.33$\pm$6.32{\%}, and 5.17$\pm$3.66{\%} over the same 
velocity range as {\cii} (Figure \ref{OIanalysis_fig}, bottom). 
No significant variation was observed in the {\oi} triplet. The {\oi} line at 
1355 {\AA} similarly does not show any significant variation between in-transit and 
out-of-transit.
Lastly, we produced time series for the flux of some stellar lines (Figure \ref{picture7ay_fig}).
The series reveals no obvious variability that could cause a false transit detection.

\section{Phenomenological model of the ion tail \label{phenomodel_appendix}}

Our nominal transit depth for {\cii} of 6.8{\%} translates into a 
projected area equivalent to a
disk of radius $R_{\rm{C\;II}}$/$R_{\rm{p}}$=15.1.
This is about the extent of {\pimenc}'s Roche lobe in the substellar direction 
($R_{\rm{L_1}}$/$R_{\rm{p}}$=13.3). 
The spectroscopic velocity of $+$10 km/s for the {\cii} ions is consistent with our photochemical-hydrodynamic predictions, 
and likely traces absorption in the vicinity of the planet's dayside  
as the gas accelerates toward the star.  
Comparable velocities are predicted by 3D models on the 
planet's dayside \citep{shaikhislamovetal2020b}
before the escaping gas interacts with the impinging stellar wind. 
Negative velocities of $-$70 km/s (and probably faster, as the stellar line becomes 
weak and the S/N poor at the corresponding wavelengths) 
suggest that the {\cii} ions are accelerated away from the star by other mechanisms such as 
tidal forces and magnetohydrodynamic interactions with the stellar wind. 
For reference, 
the velocity of the solar wind at the distance of {\pimenc} is on the order of 250 km/s
\citep{shaikhislamovetal2020b}. 
There is observational evidence for HD 209458 b, 
GJ 436 b and GJ 3470 b that their escaping atmospheres 
also result in preferential blue absorption (particularly the latter two)
 \citep{vidalmadjaretal2003,ehrenreichetal2015,bourrieretal2018}. 
Models considering the 3D geometry of the interacting stellar and planet winds 
also favor blue absorption, 
especially when the stellar wind is stronger \citep{
shaikhislamovetal2020a,shaikhislamovetal2020b} and arranges  
the escaping atoms into a tail trailing the planet. 
We consider a {\cii} tail to be realistic scenario for {\pimenc}. 
\\

\begin{figure}[h]
   \centering
   \includegraphics[angle=0,width=8.cm]{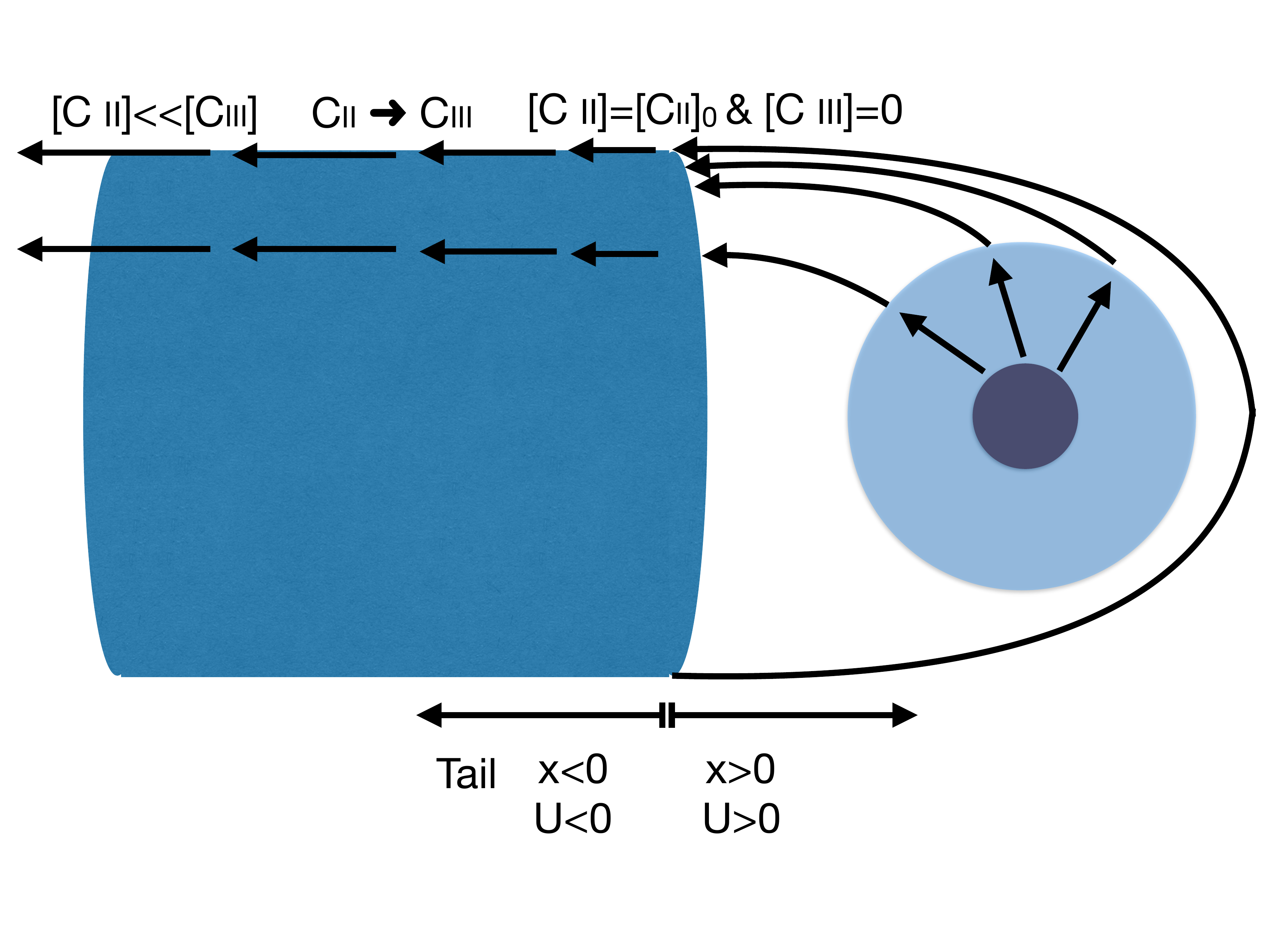}
   \includegraphics[angle=0,width=13.cm]{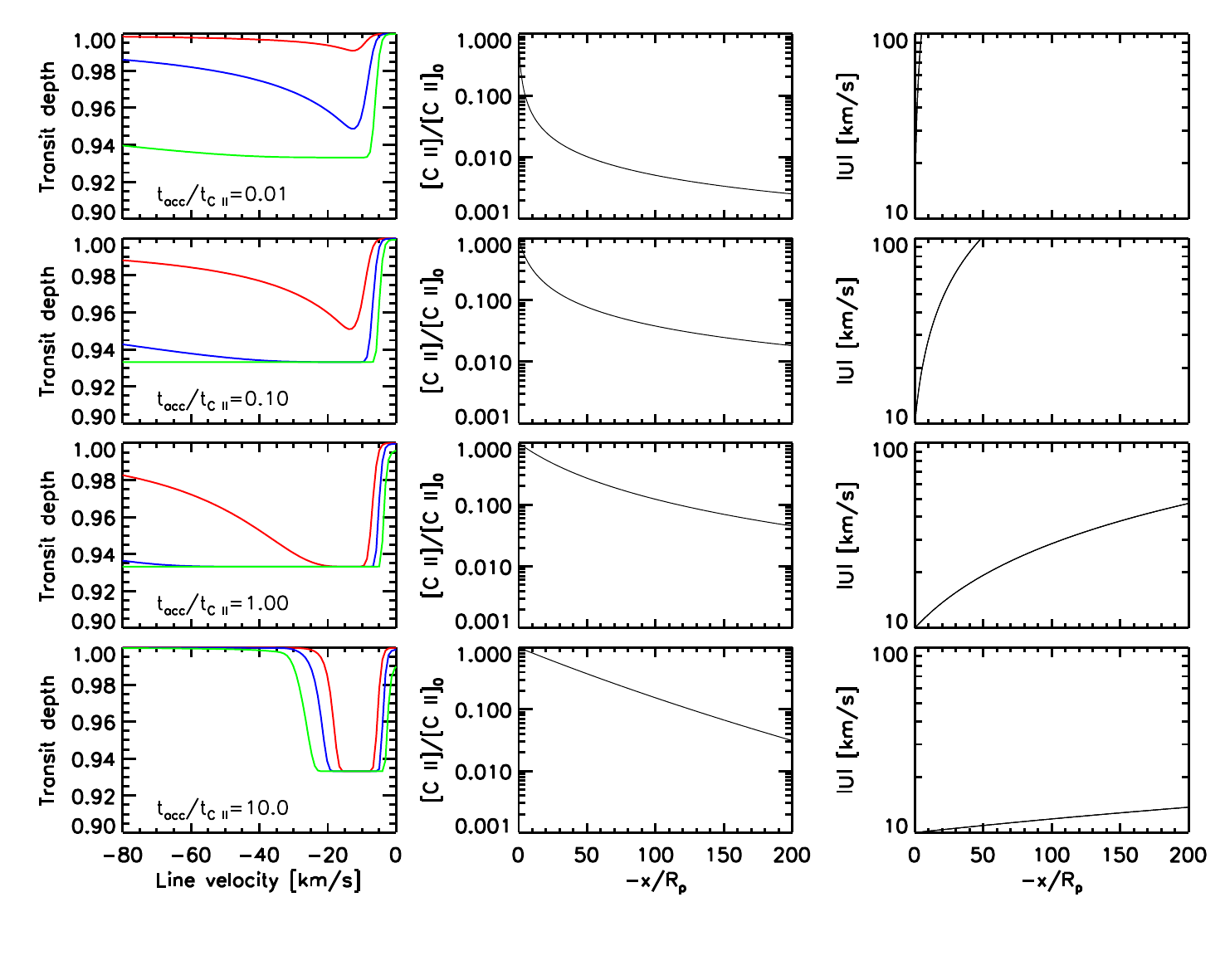}   
   \caption{\label{phenomodel_fig}
   \textbf{Top}.
   Sketch of the phenomenological model for the {\cii} tail. The ions escaping
the planet are channeled into a 
   cylindrical tail of radius 15$R_{\rm{p}}$ and accelerated. Further ionization into
   {\ciii} and the gas velocity dictate the effective length of the {\cii} tail. 
   Both $x$ (position from the tail entrance) and $U$ (velocity) are negative in the tail. 
   The star is on the right end, and the observer on the left end.
   \textbf{Bottom}.
   Examples of solution of the phenomenological model for acceleration-to-ionization 
   timescales of the {\cii} ion ranging between 0.01 and 10. 
   Left column: Transit depths for a range of {\cii} densities at the tail entry ($x$=0), 
   namely: [{\cii}]$_0$=10$^4$ (red), 10$^5$ (blue) and 10$^6$ (green) cm$^{-3}$. 
   Middle column: Ion number densities normalized to the density at $x$=0. 
   Right column: Prescribed velocity profiles.
     }
   \end{figure}

To gain intuition, we have built the phenomenological model 
of {\pimenc}'s tail sketched in Figure \ref{phenomodel_fig} (top). 
The {\cii} ions are injected into a cylindrical tail of fixed radius 15$R_{\rm{p}}$
and are subject to a prescribed velocity $U$=$-$($U_0$ $-$ $x$/$t_{\rm{acc}}$) km/s
that varies in the tail direction.  
This geometry surely simplifies the true morphology of the escaping gas, 
which is likely to resemble an opening cone tilted with respect to the star-planet line
\citep{shaikhislamovetal2020b}. This crude description allows us at the very least to 
obtain analytical expressions for some of the relevant diagnostics. 
Here, $U_0$=10 km/s (a typical value from the photochemical-hydrodynamic simulations; 
Figure \ref{panel1d_fig}) and $t_{\rm{acc}}$ is an acceleration time scale. 
(Note: $x$$<$0 in the tail, and the ions are permanently accelerating.) 
Related accelerations have been predicted by physically-motivated 3D 
models \citep{ehrenreichetal2015,shaikhislamovetal2020a,shaikhislamovetal2020b}, under
the combined effect of gravitational, inertial and radiative forces. 
Magnetic interactions with the stellar wind might additionally affect ion acceleration.
\\

The {\cii} ions photoionize further into {\ciii} with a time scale 
$t_{\rm{C\;II}}$$\sim$20 hours (calculated for unattenuated radiation from
{\pimen} at wavelengths shorter than the 508-{\AA} threshold 
and the corresponding {\cii} cross section \citep{verneryakovlev1995}). 
The collision of stellar wind particles with the planetary {\cii} ions 
might ionize further the latter (ionization potential of 24 eV), especially at the mixing layer between the 
two flows \citep{tremblinchiang2013}, but it remains unclear whether collisional ionization 
can compete on the full-tail scale with photoionization. This should be assessed in future work. 
The continuity equations that govern the model are:
\begin{eqnarray*}
\frac{d({[\rm{C}\;{\sc{II}}]}(x) U(x))}{dx}=-\frac{{[\rm{C}\;{\sc{II}}]}(x)}{t_{\rm{C\;II}}} \\
\frac{d({[\rm{C}\;{\sc{III}}]}(x) U(x))}{dx}=+\frac{{[\rm{C}\;{\sc{II}}]}(x)}{t_{\rm{C\;II}}}. 
\end{eqnarray*} 
As the flow accelerates through the tail, the total density decays and 
the {\cii} ions are converted into {\ciii}. 
The solution for the {\cii} ion number density is: 
\begin{equation*}
\frac{{[\rm{C}\;{\sc{II}}]}(x)}{{[\rm{C}\;{\sc{II}}]}_0} =
\left(
\frac{1}{1-\frac{x}{U_0 t_{\rm{acc}} }} \right) ^{1+t_{\rm{acc}}/t_{\rm{C\;II}}}.
\end{equation*}
This highly simplified exercise aims to estimate reasonable values 
for the free parameters [{\cii}]$_0$ and $t_{\rm{acc}}$ that reproduce 
the transit depth and Doppler velocities from the COS data. 
To produce the wavelength-dependent opacity, 
we integrate [{\cii}]($x$) along the line of sight keeping track of the Doppler shifts introduced 
by the varying $U$($x$). 
We represent the {\cii} cross section at rest for the 1335.7 {\AA} line by 
a Voigt lineshape with thermal ($T$=6000 K, a reference temperature in the tail; 
our findings do not depend sensitively on this temperature, 
as the absorption over a broad range of velocities is caused by the bulk velocity of the ions
rather than by their thermal broadening) and natural 
(Einstein coefficient $A_{\rm{ul}}$=2.88$\times$10$^8$ s$^{-1}$) broadening components, 
and a fractional population of the substate from which the line arises based on 
statistical weights and equal to 2/3. 
\\

Figure \ref{phenomodel_fig} (bottom) shows how [{\cii}]$_0$ and $t_{\rm{acc}}$/$t_{\rm{C\;II}}$ 
affect the absorption of the stellar line. 
Absorptions consistent with the measurements are generally found for
[{\cii}]$_0$$\ge$10$^4$ cm$^{-3}$ and $t_{\rm{acc}}$/$t_{\rm{C\;II}}$$\le$1, and
result in {\cii} tails longer than $\sim$50$R_{\rm{p}}$ and large amounts of {\cii} 
moving at $-$70 km/s. 
A key outcome of the model is that 
for $t_{\rm{acc}}$/$t_{\rm{C\;II}}$$>$1 the {\cii} ion photoionizes too quickly to 
produce significant absorption at the faster velocities. This could in principle be 
compensated with increasing amounts of {\cii} ions entering the tail, but the 
photochemical-hydrodynamic model indicates that it is not possible to go 
beyond [{\cii}]$_0$$\sim$10$^6$ cm$^{-3}$ without obtaining unrealistically large
escape rates.
For [{\cii}]$_0$$\sim$10$^4$$-$10$^6$ cm$^{-3}$, the mass fluxes of carbon atoms through the 
tail range from $\sim$10$^8$ to $\sim$10$^{10}$ g s$^{-1}$. 
These mass fluxes are comparable to the loss rates of carbon nuclei predicted by our
 photochemical-hydrodynamic model for atmospheres that
have carbon abundances larger than solar (Table \ref{summaryescape_table}).
This sets a weak constraint on the carbon abundance that 
prevents us for the time being from 
assessing whether carbon is a major constituent of {\pimenc}'s atmosphere. 
\\

This insight into {\pimenc}'s carbon abundance, even though of limited diagnostic 
value, is consistent with the interpretation of in-transit absorption by {\cii} 
at the hot Jupiter HD 209458 b \citep{vidalmadjaretal2004,benjaffelsonahosseini2010,linskyetal2010}.
Both HD 209458 b and {\pimenc} transit Sun-like stars and exhibit similar transit depths. 
3D models of HD 209458 b \citep{shaikhislamovetal2020a} show 
that its {\cii} absorption signal is explained by carbon in solar abundance. 
Because the predicted mass loss rate of HD 209458 b is an order of magnitude higher than 
for {\pimenc} (a result mainly from its lower density), a reasonable guess is that a
$\times$10 solar enrichment for {\pimenc} will result in comparable transit depths.
Future 3D modeling of
the interaction of the {\cii} ions with the stellar wind
and the planet's magnetic field lines should help refine these conclusions.

\section{Photochemical-hydrodynamic model \label{hydromodel_appendix}}

We investigated {\pimenc}'s extended atmosphere
with a 1D photochemical-hydrodynamic model that 
solves the gas equations at pressures $\le$1 $\mu$bar \citep{garciamunozetal2020}.
Heating occurs from absorption of stellar photons by the atmospheric neutrals. 
Cooling is parameterized as described in our previous work, and 
includes emission from {\hthreep} in the infrared, 
{\lalpha} in the FUV, atomic oxygen at 63 and 147 $\mu$m, and rotational cooling from 
H$_2$O, OH and CO also in the infrared. 
We adopted a NLTE formulation of {\hthreep} cooling that
captures the departure of the ion's population from equilibrium at low {\htwo} densities.
We included a parameterization of {\cotwo} cooling at 15 $\mu$m \citep{gordietsetal1982}. 
The model considers 26 chemicals 
(H$_2$, H, He, CO$_2$, CO, C, H$_2$O, OH, O, O$_2$, 
H$_2^+$, H$^+$, H$_3^+$, He$^+$, HeH$^+$, CO$_2^+$, CO$^+$, C$^+$, HCO$_2^+$, HCO$^+$, 
H$_2$O$^+$, H$_3$O$^+$, OH$^+$, O$^+$, O$_2^+$ and electrons) that
participate in 154 chemical processes. 
It does not include hydrocarbon chemistry, 
although this omission is not important as hydrocarbons are
rapidly lost at low pressures in favor of other carbon-bearing species
\citep{mosesetal2013}.
\\

To explore a broad range of compositions, we adopted nuclei fractions
$Y_{\rm{C}^*}$ and $Y_{\rm{O}^*}$ such that 
$Y_{\rm{C}^*}$+$Y_{\rm{O}^*}$ goes from a few times 10$^{-6}$ to $\sim$0.4, 
and $Y_{\rm{C}^*}$/$Y_{\rm{O}^*}$ ratios from 0.1 to 10. 
We imposed $Y_{\rm{He}^*}$=0.1$Y_{\rm{H}^*}$.
By definition $Y_{\rm{H}^*}$+$Y_{\rm{He}^*}$+$Y_{\rm{C}^*}$+$Y_{\rm{O}^*}$=1, and
thus the composition is specified by only two nuclei fractions.
With the above information, we estimated the corresponding vmrs from Figure 7 of \citet{huseager2014}
and assigned them as boundary conditions of our extended atmosphere model.
For the other gases, we adopted zero vmrs.
Hydrocarbons (e.g. C$_2$H$_2$, CH$_4$) become abundant in the bulk atmosphere
for $Y_{\rm{C}^*}$/$Y_{\rm{O}^*}$$\geq$2. 
Because our model does not currently include hydrocarbons, 
we transferred all the C nuclei at the base of the extended atmosphere
from C$_2$H$_2$ and CH$_4$ into CO and C. 
\\

Table \ref{summaryescape_table} summarizes 30 cases, each 
for a different bulk atmospheric composition. 
For cases 05-06 and 11-12 we \textcolor{black}{first} assumed for their vmrs in the bulk atmosphere: 
\begin{itemize}
\item
(05) {\htwo}: 7.66$\times$10$^{-1}$; 
         He: 1.66$\times$10$^{-1}$; 
         CO: 6.90$\times$10$^{-3}$; 
         {\htwoo}: 6.21$\times$10$^{-2}$. 
\item
(06) 
He: 1.46$\times$10$^{-1}$; \cotwo: 9.76$\times$10$^{-2}$; {\htwoo}: 7.32$\times$10$^{-1}$; 
{\otwo}: 2.44$\times$10$^{-2}$.
\item
(11) 
{\htwo}: 7.88$\times$10$^{-1}$; 
He: 1.63$\times$10$^{-1}$; 
CO: 2.48$\times$10$^{-2}$; 
\htwoo: 2.48$\times$10$^{-2}$. 
\item 
(12)  
{\htwo}: 4.47$\times$10$^{-1}$; 
He: 1.18$\times$10$^{-1}$; 
\cotwo: 1.45$\times$10$^{-1}$; 
CO: 1.45$\times$10$^{-1}$; 
\htwoo: 1.45$\times$10$^{-1}$. 
\end{itemize}
However, our numerical experiments 
showed that H$_2$O (but also H$_2$ and O$_2$) were unstable for these cases, 
and their number densities dropped by orders of magnitude in a few elements of the spatial grid.
This is evidence that these molecules chemically transform before reaching the $\mu$bar pressure level. 
To avoid numerical difficulties, in these four cases 
we prescribed the vmrs at the $\mu$bar pressure level by replacing the unstable molecules
by their atomic constituents while preserving the original nuclei fractions.
They are indicated with a $\dagger$ in Table \ref{summaryescape_table}, 
which lists the adopted vmrs. 
\\

The mass fraction  of e.g. {\htwo} and He
is given by the ratio vmr(\htwo)$\;\mu_{\rm{H}_2}$/($\sum_{s}$vmr$_s$$\;\mu_s$)
and vmr(He)$\;\mu_{\rm{He}}$/($\sum_{s}$vmr$_s$$\;\mu_s$), 
with the summation extending over all species, respectively.
The mass fraction of heavy volatiles in the bulk atmosphere $Z$ is calculated as 
one minus the added mass fractions of {\htwo} and He.
\\

The right hand side of Table \ref{summaryescape_table} summarizes the model outputs. 
$\dot{m}_{\rm{H}}$($@$\textit{Mach}=1) quotes the mass flux of neutral 
{\hi} atoms at the sonic point. It departs from 
the mass loss rate of H nuclei ($\dot{m}_{\rm{H}^*}$) if 
at the sonic point hydrogen is ionized. 
$\dot{m}_{\rm{He}^*}$, $\dot{m}_{\rm{C}^*}$, $\dot{m}_{\rm{O}^*}$ and $\dot{m}$
quote the mass loss rates of the specified nuclei and the net mass loss rate of the 
atmosphere. All mass loss rates are calculated over a 4$\pi$ solid angle.
\\

Table \ref{summaryescape_table} shows that $\dot{m}_{\rm{H}}$($@$\textit{Mach}=1) 
is about 5$\times$10$^{9}$ g s$^{-1}$ for a {\htwo}/He atmosphere. 
3D simulations of {\pimenc}'s extended atmosphere \citep{shaikhislamovetal2020b} show that the {\lalpha}
absorption signal varies monotonically with the mass flux of protons in the 
stellar wind. According to their numerical experiments (their Figure 4), 
increases in the stellar wind flux by a factor of a few result in deeper transits
by a similar factor of a few. 
Because ENA generation depends on both the flux of protons in the stellar wind and the 
flux of neutral hydrogen in the planet wind, we estimate that a factor of a few 
(we take $\times$1/4) decrease in $\dot{m}_{\rm{H}}$($@$\textit{Mach}=1) with respect
to the case of an {\htwo}/He atmosphere suffices to bring ENA generation to undetectable
levels for HST/STIS. We take $\dot{m}_{\rm{H}}$($@$\textit{Mach}=1)=1.25$\times$10$^{9}$ g s$^{-1}$
as our approximate threshold for non-detection of {\lalpha} absorption. 
A refined estimate of this threshold calls for 3D simulations over a variety of 
bulk atmospheric compositions.

\section{Interior structure model \label{interior_appendix}}

As the adopted intrinsic temperature affects the predicted atmospheric
mass \textcolor{black}{(because of its impact on the scale height)}, we constrain its possible values through thermal evolution calculations. 
For an adiabatic planetary envelope that cools efficiently by convection 
under the moderating effects of atmospheric opacity, a 
pattern emerges: The lower \textcolor{black}{the present-day} $T_{\rm int}$ and the more massive 
the atmosphere is, the longer it takes to cool down to that state from an initial hot state
\textcolor{black}{($T_{\rm int}$ much larger than 100 K)}
after formation (time $t$=0). 
\\

\begin{figure}[ht]
\includegraphics[width=8cm]{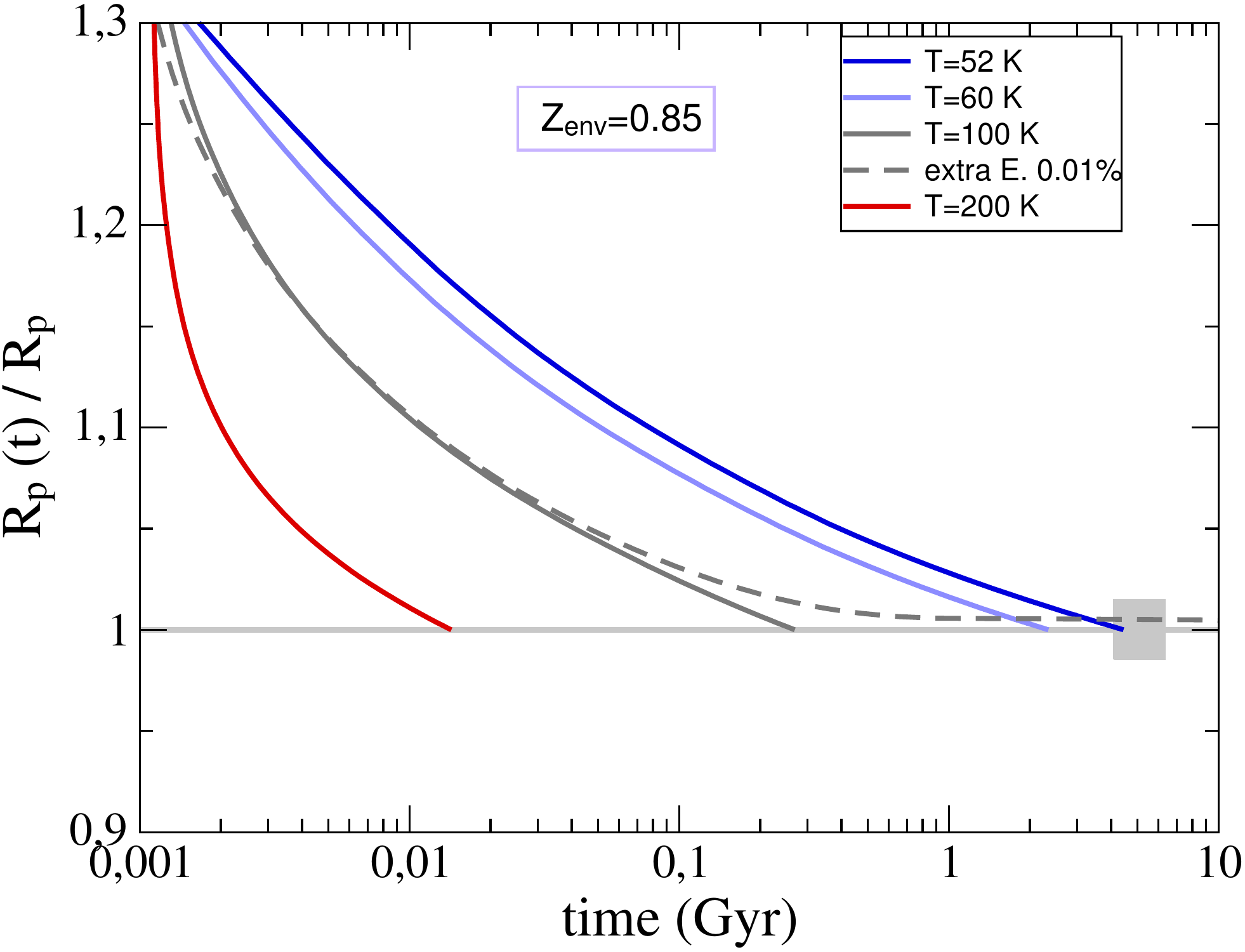}\includegraphics[width=8cm]{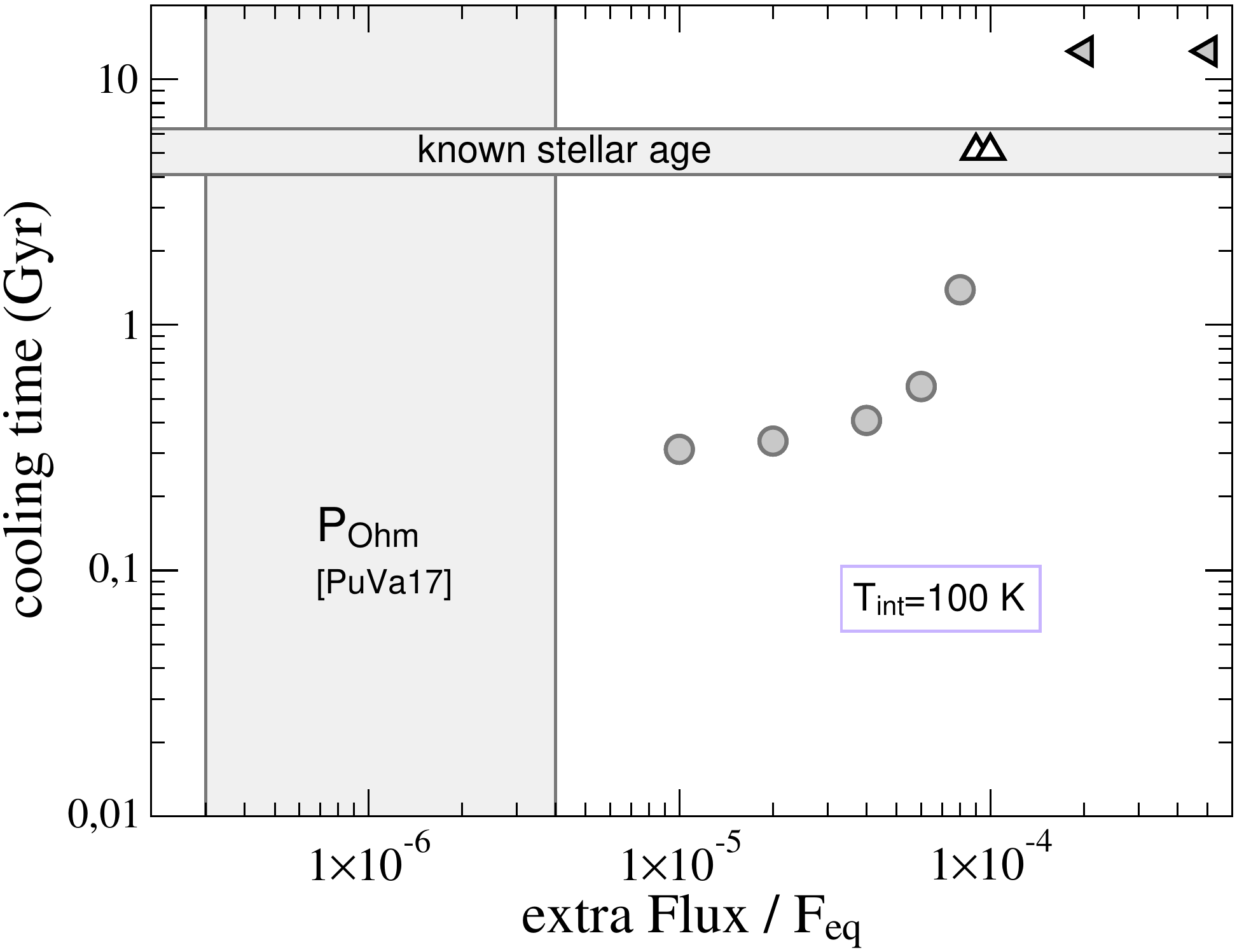}
\includegraphics[width=8cm]{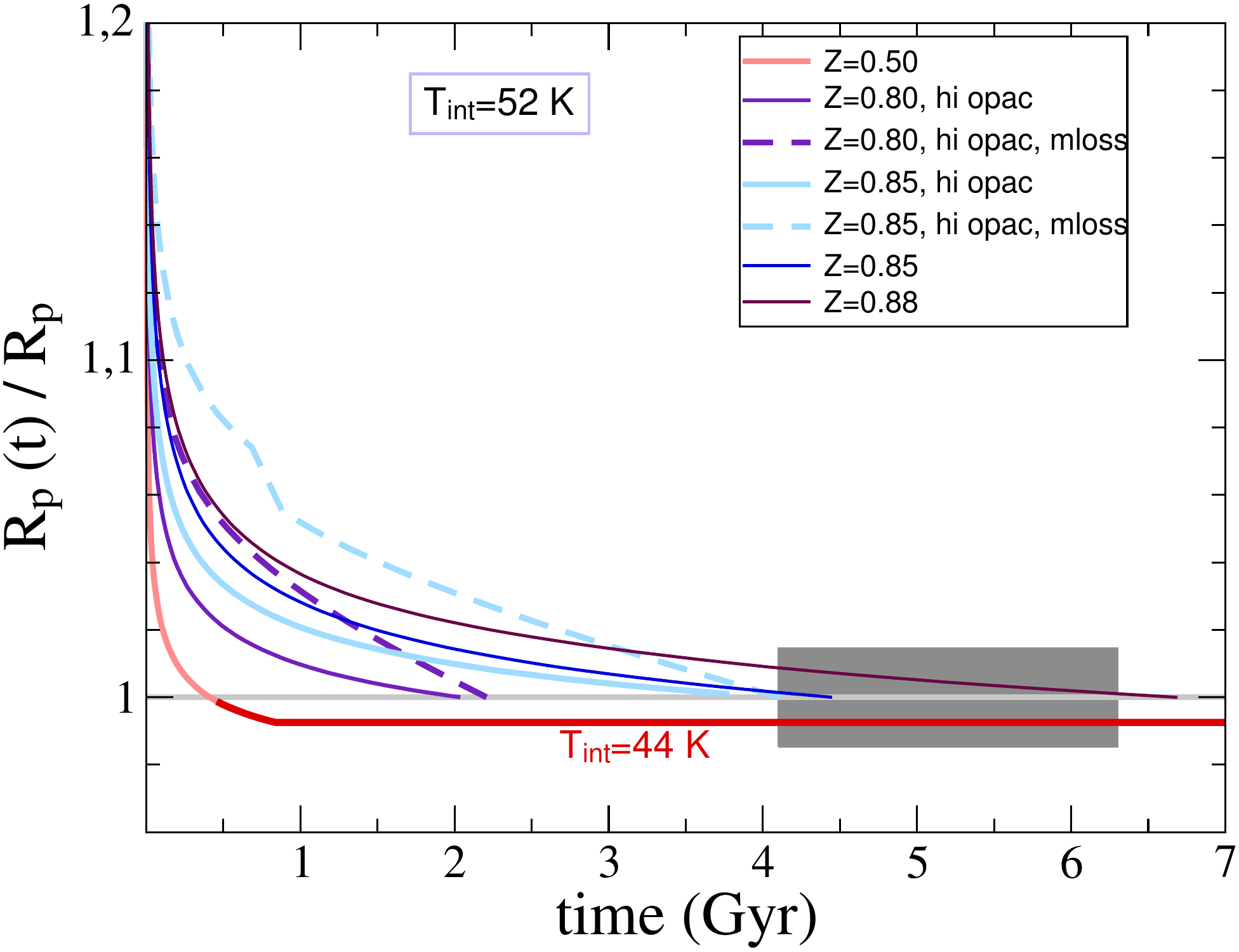}\includegraphics[width=8cm]{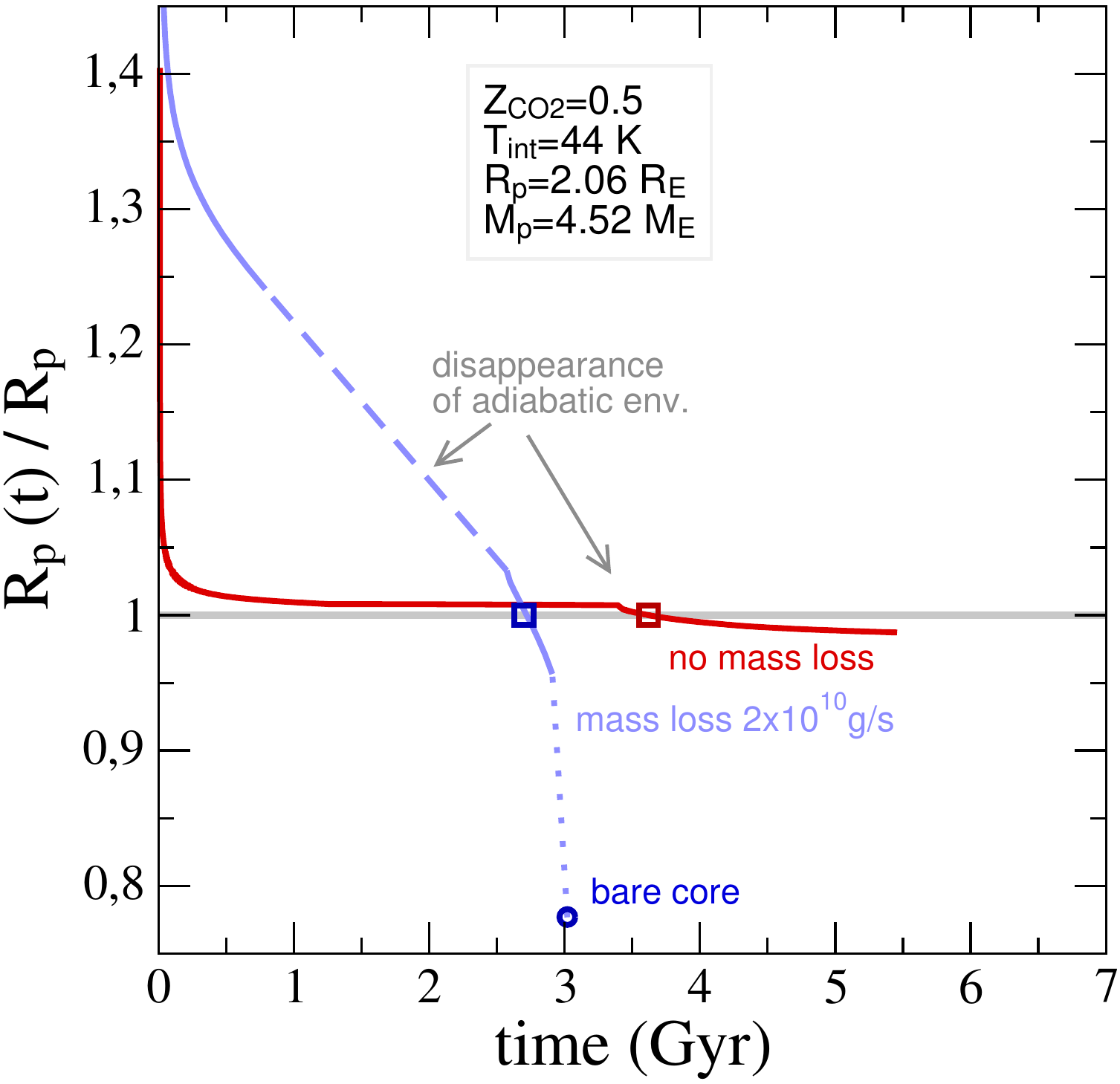}
\caption{\label{planetinterior_fig}
}
\end{figure}

Figure \ref{planetinterior_fig} (top left) shows evolution tracks for 
different \textcolor{black}{present-day} $T_{\rm int}$ and a heavy volatile mass fraction 
$Z$=0.85.
The largest $T_{\rm int}$ that yields a cooling time in agreement with the system's age
is 52~K, although $T_{\rm int}$=60 K might yield a solution 
where the radius is matched within the measurement uncertainties. 
Assuming $T_{\rm int}$=100 K requires an extra heating source that maintains such a high heat flux 
at present times. For inflated hot Jupiters, extra heating of $\sim$0.1{\%} to a few percent of the 
incident stellar irradiation is required to explain their large radius, which is consistent with Ohmic heating 
\citep{thorngrenfortney2018}.  
For {\pimenc}, less extra heating  is required
($\sim$0.01$\%$; gray dashed curve). However, the mechanism that may provide this extra heating to 
the smaller {\pimenc} is not obvious. 
While Ohmic heating may occur at sub-Neptunes \citep{puvalencia2017},
the predicted amount falls short by two orders of magnitude of what is needed for {\pimenc}  
(Figure \ref{planetinterior_fig}, top right). 
Another option is tidal heating provided that the planet is on an eccentric orbit, although 
the orbital eccentricity remains poorly constrained 
\citep{gandolfietal2018,huangetal2018,damassoetal2020}. 
For an eccentricity $e$=$0.001$ and tidal quality factor $Q_{\rm{p}}$$\geq$10$^3$ we find 
that tidal heating affects negligibly the planetary cooling.
Tidal heating however becomes effective at extending the cooling time of an atmosphere with 
$T_{\rm int}$=100 K up to the current system's age if, for example, $e$=0.02 and 
$Q_{\rm{p}}$$\sim$10$^3$ (about three times the Earth's tidal quality factor), 
or if $e$=0.1 and $Q_{\rm{p}}$$\sim$5$\times$10$^4$ (about the Saturn/Uranus/Neptune value). 
Further, we estimated the circularization time for these two configurations 
\citep{jacksonetal2008} to be $\tau_{\rm{circ}}$=0.6 and 30 Gyr respectively. 
Even the shortest of them is on the order of our prescribed survival time. It is thus 
reasonable to expect that the recent history of {\pimenc}'s atmosphere may have occurred
while the planet followed an eccentric orbit that could have sustained $T_{\rm int}$$\sim$100 K.
It is also conceivable that the outer companion in the {\pimen} planetary system may 
endow a non-negligible eccentricity to the innermost planet's orbit 
\citep{derosaetal2020,kunovachodzicetal2020,xuanwyatt2020}, as is found in some 
close-in sub-Neptune-plus-cold Jupiter systems. 
Importantly, to the effects of planet mass loss, 
higher $T_{\rm int}$ values that imply 
lower $M_{\rm atm}$ are less likely to survive over time scales of Gigayears, which sets
a limit to how high $T_{\rm int}$ can be given that {\pimenc} still hosts an atmosphere. 
Finally, we adopt $T_{\rm int}$=100 K as a reasonable upper limit for {\pimenc}. 
\\

Figure \ref{planetinterior_fig} (bottom left) shows evolution tracks for $Z$ from 0.5 to 0.88
at \textcolor{black}{present-day} $T_{\rm int}$=52 K. The lower the adopted $Z$, the less massive the atmosphere is 
and the quicker it cools down and contracts. 
Eventually, the planet adopts a state of \textcolor{black}{(nearly)} equilibrium evolution with the incident flux, where contraction
slows down and further cooling progresses on very long timescales. 
In this state, the planet radiates ($\sim$(100/1150)$^4$$\sim$6$\times$10$^{-5}$)
only 0.006{\%} more than if it was in true equilibrium with the stellar irradiation. 
Mass fractions $Z$$<$0.5 may be possible for present $T_{\rm int}$$<$44 K. 
For comparison, Saturn has $T_{\rm int}$$\sim$77 K and Neptune $\sim$53 K, and thus we 
do not expect that $T_{\rm int}$ at {\pimenc} will be much higher than for them
given that its mass is much lower. 
Also, a non-zero eccentricity may keep $T_{\rm int}$ above such values. 
Mass loss, fixed in the preparation of Figure \ref{planetinterior_fig} (bottom left) to 
2$\times$10$^{10}$ g/s, prolongs the cooling time immediately after formation 
but speeds up the contraction well into the future as the planet loses its envelope. 
\\

We have simulated the future evolution of {\pimenc} for some of the compositions
deemed realistic for present-day {\pimenc} with the goal of exploring whether the planet 
will ever cross the radius valley. 
As an example, Figure \ref{planetinterior_fig} (bottom right) shows that for $Z$({\cotwo})=0.50 and
$T_{\rm{int}}$=44 K, {\pimenc} will turn into a bare rocky core in about 0.5 Gyr.
The same conclusion is found for other $Z$--$T_{\rm{int}}$ combinations.
\\

The interior structure model calculates the atmospheric pressure-temperature ($p$--$T$) 
profile using an analytical formulation \citep{guillot2010} that 
depends on the ratio $\gamma$=$\kappa_{\rm vis}$/$\kappa_{\rm IR}$ of
constant visible and IR opacities in the $T$--optical depth relation and on the local
opacity. For the IR opacities, we take the 
Rosseland mean $\kappa_{\rm R}$ for solar composition \citep{freedmanetal2008}
as our baseline.
In our opacity 1 setting, we adopt $\gamma$=0.123 and 
$\kappa_{\rm IR}$=1$\times$$<$$\kappa_{\rm R}$$>$ (bracket $<$$>$ denotes average over wavelength) 
which was confirmed to reproduce well
the $p$--$T$ profiles published for the H$_2$/He/H$_2$O atmosphere of K2-18 \citep{scheucheretal2020}.
As {\pimenc}'s atmosphere may also contain large abundances of {\cotwo}, 
our opacity 2 setting adopts $\gamma$=0.500 and $\kappa_{\rm IR}$=2$\times$$<$$\kappa_{\rm R}$$>$, 
which is appropriate for a CO$_2$-dominated atmosphere with some admixture of H$_2$/He.
Using two sets of opacities for each interior model calculation is a pragmatic
way of bracketing the real opacity of {\pimenc}'s atmosphere. In a conservative spirit, 
our constraints on the planet's bulk composition from the atmospheric stability argument 
utilize the opacity setting that results in the lowest $Z$.

\clearpage
\newpage

\textbf{Caption for Table} \ref{summaryescape_table}. 
\\

Summary of photochemical-hydrodynamic model runs. 
Columns 2--4 quote the assumed nuclei fractions for H, C and O in the bulk atmosphere.  
Columns 6--13 quote the adopted vmrs at the $\mu$bar pressure level 
for H$_2$, H, He, CO$_2$, CO, C, H$_2$O and O, and 14 quotes the corresponding heavy mass fraction. 
Column 15 quotes the mass flux of {\hi} atoms at the location of the sonic point, 
which occurs for all cases within the range $r_{{Mach}=1}$/$R_{\rm{p}}$=3.5--4.6.
Columns 16--20 quote the mass loss rates of H, He, C and O nuclei, and  21 quotes 
the net mass loss rate. 
$\dagger$: For these cases, we specified the vmrs at 1 $\mu$bar
considering that molecules such as H$_2$, H$_2$O and O$_2$ become photochemically unstable before reaching the $\mu$bar pressure level.
We also assume that the 1-$\mu$bar pressure level occurs at the radial
distance $\sim$2.06$R_{\Earth}$
for the TESS radius of the planet. This neglects the vertical extent of the
region between a few mbars and 1 $\mu$bar, a reasonable approximation for moderate
atmospheric temperatures. 
We impose that at the 1-$\mu$bar level
the temperature coincides with the planet's $T_{\rm{eq}}$ (1150 K).

\newpage 

\begin{sidewaystable}

\fontsize{6}{10}\selectfont

    \centering
\caption{}             
\label{summaryescape_table}      
\centering                          

\begin{tabular}{|p{0.30cm}p{0.89cm}p{0.89cm}p{0.89cm}p{0.55cm}p{0.76cm}p{0.76cm}p{0.76cm}p{0.76cm}p{0.76cm}p{0.76cm}p{0.76cm}p{0.76cm}p{0.76cm}|p{1.16cm}p{0.45cm}p{0.77cm}p{0.77cm}p{0.77cm}p{0.70cm}|}

\hline                 

Case & \centering{$Y_{\rm{H}^*}$} & \centering{$Y_{\rm{C}^*}$} & \centering{$Y_{\rm{O}^*}$} & \centering{$Y_{\rm{C}^*}$/$Y_{\rm{O}^*}$} & \multicolumn{8}{c}{Volume mixing ratio at $R_{\rm{1\mu bar}}$} &  & $\dot{m}_{\rm{H}}$($@$$M$=1) 
& $\dot{m}_{\rm{H}^*}$ & $\dot{m}_{\rm{He}^*}$ & $\dot{m}_{\rm{C}^*}$ & $\dot{m}_{\rm{O}^*}$ & $\dot{m}_{\rm{}}$ \\
     &                &                &  &  & \centering{{\htwo}} & \centering{H} & \centering{He} & \centering{{\cotwo}} & \centering{CO} & \centering{C} & \centering{{\htwoo}} & \centering{O} & $Z$ &[10$^9$gs$^{-1}$] &  \multicolumn{5}{c|}{[10$^9$gs$^{-1}$]} \\

\hline                        

   01  & 9.09(-1) & 3.64(-7) & 3.64(-6)         & 0.1 & 8.3(-1) & 0.0     & 1.7(-1) & 0.0 & 6.7(-7) & 0.0 & 6.0(-6) & 0.0 & 5.2(-5) &4.4  & 6.9 & 2.7 & 3.0(-5) & 4.1(-4) & 9.6 \\
   02  & 9.09(-1) & 3.64(-6) & 3.64(-5)         & 0.1 & 8.3(-1) & 0.0     & 1.7(-1) & 0.0 & 6.7(-6) & 0.0 & 6.0(-5) & 0.0 & 5.2(-4) &6.7  & 9.7 & 3.8 & 4.3(-4) & 5.9(-3) & 13.5 \\
   03  & 9.09(-1) & 3.64(-5) & 3.64(-4)         & 0.1 & 8.3(-1) & 0.0     & 1.7(-1) & 0.0 & 6.7(-5) & 0.0 & 6.0(-4) & 0.0 & 5.2(-3) &11.9  & 15.0 & 5.9 & 6.9(-3) & 9.3(-2) & 21.0 \\
   04  & 9.06(-1) & 3.64(-4) & 3.64(-3)         & 0.1 & 8.3(-1) & 0.0     & 1.7(-1) & 0.0 & 6.7(-4) & 0.0 & 6.0(-3) & 0.0 & 5.2(-2) &15.3  & 18.9 & 7.3 & 8.5(-2) & 1.3 & 27.4 \\ 

   05$\dagger$ & 8.73(-1) & 3.64(-3) & 3.64(-2) & 0.1 & 0.0     & 8.8(-1) & 8.8(-2) & 3.7(-3) & 0.0 & 0.0 & 0.0  &  2.9(-2) & 3.7(-1)&8.4 & 11.5 & 4.5 & 5.9(-1)  & 7.5 & 24.1 \\

   06$\dagger$ & 5.45(-1) & 3.64(-2) & 3.64(-1) & 0.1 & 0.0     & 5.9(-1) & 5.9(-2) & 3.9(-2) & 0.0 & 0.0 & 0.0  &  3.1(-1) & 9.7(-1) &0.35 & 1.4 & 4.8(-1) & 8.2(-1)  & 11.7 & 14.4\\

\hline
\hline

   07 & 9.09(-1) & 1.33(-6) & 2.67(-6)          & 0.5 & 8.3(-1) & 0.0     & 1.7(-1) & 0.0 &  2.4(-6) & 0.0 & 2.4(-6) & 0.0 & 4.8(-5)&4.4 &  6.8  & 2.7 & 1.1(-4) & 2.9(-4) &9.5 \\
   08 & 9.09(-1) & 1.33(-5) & 2.67(-5)          & 0.5 & 8.3(-1) & 0.0     & 1.7(-1) & 0.0 &  2.4(-5) & 0.0 & 2.4(-5) & 0.0 & 4.8(-4)&5.6 &  8.3  & 3.3 & 1.4(-3) & 3.6(-3) &11.6 \\
   09 & 9.09(-1) & 1.33(-4) & 2.67(-4)          & 0.5 & 8.3(-1) & 0.0     & 1.7(-1) & 0.0 &  2.4(-4) & 0.0 & 2.4(-4) & 0.0 & 4.8(-3)&11.6 &  14.9 & 5.9 & 2.5(-2) & 6.7(-2) &20.9 \\
   10 & 9.06(-1) & 1.33(-3) & 2.67(-3)          & 0.5 & 8.3(-1) & 0.0     & 1.7(-1) & 0.0 &  2.4(-3) & 0.0 & 2.4(-3) & 0.0 & 4.6(-2)&12.8 &  15.9 & 6.2 & 2.7(-1)  & 7.1(-1)  &23.2 \\   
   11$\dagger$ & 8.73(-1) & 1.33(-2) & 2.67(-2) & 0.5 & 8.1(-1) & 0.0     & 1.6(-1) & 2.5(-2) & 0.0  & 0.0 & 0.0     & 0.0 & 3.4(-1) &9.3 &  12.5 & 4.5 & 1.9  & 5.4  &23.3\\    
   12$\dagger$& 5.45(-1) & 1.33(-1) & 2.67(-1) & 0.5 & 0.0      & 7.4(-1)& 7.4(-2) & 1.8(-1) & 0.0  & 0.0 & 0.0     & 0.0 & 9.0(-1) &0.54 &  1.7 & 5.7(-1) & 3.5  & 9.3 & 14.6 \\ 

\hline
\hline

   13 & 9.09(-1) & 2.00(-6) & 2.00(-6)          & 1 & 8.3(-1) & 0.0     & 1.7(-1) & 0.0 &  3.7(-6) & 0.0 & 0.0 & 0.0 & 4.4(-5)&4.4 &  6.8  & 2.7 & 1.7(-4) & 2.2(-4) &9.5\\
   14 & 9.09(-1) & 2.00(-5) & 2.00(-5)          & 1 & 8.3(-1) & 0.0     & 1.7(-1) & 0.0 &  3.7(-5) & 0.0 & 0.0 & 0.0 & 4.4(-4)&4.8 &  7.3  & 2.8 & 1.8(-3) &  2.3(-3) &10.1\\
   15 & 9.09(-1) & 2.00(-4) & 2.00(-4)          & 1 & 8.3(-1) & 0.0     & 1.7(-1) & 0.0 &  3.7(-4) & 0.0 & 0.0 & 0.0 & 4.4(-3)&9.0 &  12.3  & 4.8 & 3.1(-2) &  4.1(-2) &17.2\\
   16 & 9.05(-1) & 2.00(-3) & 2.00(-3)          & 1 & 8.3(-1) & 0.0     & 1.7(-1) & 0.0 &  3.7(-3) & 0.0 & 0.0 & 0.0 & 4.2(-2)&10.7 &  13.3  & 5.3 & 3.4(-1) &  4.5(-1) &19.4\\
   17 & 8.73(-1) & 2.00(-2) & 2.00(-2)          & 1 & 8.0(-1) & 0.0     & 1.6(-1) & 0.0 &  3.7(-2) & 0.0 & 0.0 & 0.0 & 3.1(-1)&4.9 &  6.8   & 2.6 & 1.7 &  2.3 & 13.4 \\
   18 & 5.45(-1) & 2.00(-1) & 2.00(-1)          & 1 & 5.2(-1) & 0.0     & 1.0(-1) & 0.0 &  3.8(-1) & 0.0 & 0.0 & 0.0 & 8.8(-1)&0.45 &  1.3   & 0.4 & 3.7 &  4.9 & 10.3 \\
   
\hline
\hline

   19 & 9.09(-1) & 2.67(-6) & 1.33(-6)          & 2 & 8.3(-1) & 0.0     & 1.7(-1) & 0.0 &  2.4(-6) & 2.4(-6) & 0.0 & 0.0 & 4.2(-5)&4.7 &  7.3  & 2.9 & 2.4(-4) & 1.6(-4) & 10.2\\
   20 & 9.09(-1) & 2.67(-5) & 1.33(-5)          & 2 & 8.3(-1) & 0.0     & 1.7(-1) & 0.0 &  2.4(-5) & 2.4(-5) & 0.0 & 0.0 & 4.2(-4)&4.6 &  7.1  & 2.8 & 2.3(-3) & 1.5(-3) & 9.9\\
   21 & 9.09(-1) & 2.67(-4) & 1.33(-4)          & 2 & 8.3(-1) & 0.0     & 1.7(-1) & 0.0 &  2.4(-4) & 2.4(-4) & 0.0 & 0.0 & 4.2(-3)&4.7 &  7.2  & 2.8 & 2.4(-2) & 1.5(-2) & 10.0 \\
   22 & 9.05(-1) & 2.67(-3) & 1.33(-3)          & 2 & 8.3(-1) & 0.0     & 1.7(-1) & 0.0 &  2.4(-3) & 2.4(-3) & 0.0 & 0.0 & 4.0(-2)&5.2 &  7.2  & 2.8 & 2.4(-1) & 1.5(-1) & 10.4 \\
   23 & 8.73(-1) & 2.67(-2) & 1.33(-2)          & 2 & 7.9(-1) & 0.0     & 1.6(-1) & 0.0 &  2.4(-2) & 2.4(-2) & 0.0 & 0.0 & 3.0(-1)&4.4 &  5.4  & 2.1 & 1.8 & 1.2 & 10.5 \\
   24 & 5.45(-1) & 2.67(-1) & 1.33(-1)          & 2 & 4.6(-1) & 0.0     & 9.2(-2) & 0.0 &  2.2(-1) & 2.2(-1) & 0.0 & 0.0 & 8.7(-1) &0.48 &  1.3  & 0.5 & 6.1 & 3.9 & 11.8 \\
   
\hline   
   25 & 9.09(-1) & 3.64(-6) & 3.64(-7)          & 10 & 8.3(-1) & 0.0     & 1.7(-1) & 0.0 &  6.7(-7) & 6.0(-6) & 0.0 & 0.0 & 3.9(-5)&4.8 &  7.5  & 2.9 & 3.4(-4) & 4.4(-5) & 10.4 \\
   26 & 9.09(-1) & 3.64(-5) & 3.64(-6)          & 10 & 8.3(-1) & 0.0     & 1.7(-1) & 0.0 &  6.7(-6) & 6.0(-5) & 0.0 & 0.0 & 3.9(-4)&4.8 &  7.5  & 3.0 & 3.4(-3) & 4.4(-4) & 10.5 \\
   27 & 9.09(-1) & 3.64(-4) & 3.64(-5)          & 10 & 8.3(-1) & 0.0     & 1.7(-1) & 0.0 &  6.7(-5) & 6.0(-4) & 0.0 & 0.0 & 3.9(-3)&5.0 &  7.8  & 3.1 & 3.6(-2) & 4.6(-3) & 10.9 \\
   28 & 9.05(-1) & 3.64(-3) & 3.64(-4)          & 10 & 8.3(-1) & 0.0     & 1.7(-1) & 0.0 &  6.7(-4) & 6.0(-3) & 0.0 & 0.0 & 3.7(-2)&5.3 &  7.9  & 3.1 & 3.7(-1) & 4.7(-2) & 11.4 \\
   29 & 8.73(-1) & 3.64(-2) & 3.64(-3)          & 10 & 7.8(-1) & 0.0     & 1.6(-1) & 0.0 &  6.5(-3) & 5.8(-2) & 0.0 & 0.0 & 2.9(-1)&5.1 &  6.6  & 2.6 & 3.1 & 4.0(-1) & 12.7 \\
   30 & 5.46(-1) & 3.64(-1) & 3.64(-2)          & 10 & 3.9(-1) & 0.0     & 7.9(-2) & 0.0 &  5.3(-2) & 4.7(-1) & 0.0 & 0.0 & 8.7(-1)&0.67 &  1.5  & 5.9(-1) & 11.1 & 1.4 & 14.6 \\

\hline

\end{tabular}
\end{sidewaystable}

\clearpage
\newpage

\clearpage
\newpage

\clearpage
\newpage

\clearpage
\newpage

\textbf{Caption for Figure} \ref{planetinterior_fig}. 

\textbf{Top Left}.
Temporal evolution of planet radius $R_{\rm{p}}$($t$) scaled by the measured $R_{\rm{p}}$=2.06$R_{\Earth}$ 
for models with $Z$({\cotwo})=0.85 and present-day $T_{\rm{int}}$ values as labeled. 
Both $R_{\rm{p}}$ and $T_{\rm{int}}$ decrease with time as the planet cools.
The gray box indicates the uncertainty in present radius ($\pm$0.03$R_{\Earth}$) 
and stellar age (5.2$\pm$1.1 Gyr).
For the gray dashed curve of $T_{\rm{int}}=100$ K, the evolution calculations 
include an extra energy source that converts 0.01\% of the incident energy
flux into heating of the interior. 
\textbf{Top Right}.
Cooling time, defined as the time to reach the measured planet radius $R_{\rm{p}}$. 
Symbols are placed at 5.1 Gyr (13 Gyr) if contraction is halted, i.e. if $R_p$ becomes independent 
of time and agrees with (remains larger than) the measured $R_p$.  
All runs are for interior models with $T_{\rm{int}}$=100 K, $Z$=0.85, and no mass loss.
\textbf{Bottom Left}. 
Same as top left panel but for models with a lower $T_{\rm int}$=52 K and $Z$ values as labeled.
The moderate $Z$=0.50 model seems to fall short in cooling time. However, evolving that model further to 
$T_{\rm int}$=44 K leads to equilibrium evolution and $R_{\rm{p}}$(t) still in agreement with the measured value.
\textbf{Bottom Right}. 
Planet size evolution considering a fiducial mass loss 
of 2$\times$10$^{10}$ g s$^{-1}$ (blue) and omitting mass loss (red) for $Z$({\cotwo})=0.5 
and $T_{\rm{int}}$=44 K. 
Equilibrium is reached when mass loss is omitted. When mass loss is taken into account, 
the planet continues shrinking as it sheds its atmosphere until it turns into a bare rocky core, 
which is predicted to happen within 0.5 Gyr from now.
The squares indicate potential present states of {\pimenc}, from where the evolution is
calculated backward and forward in time.

\newpage

\end{document}